\documentclass{jpp}

\usepackage{amsmath}				
\usepackage{amssymb}				
\usepackage{mathrsfs}					
\usepackage[mathscr]{euscript}	
\usepackage{xspace}					
\usepackage{graphicx}
\usepackage{braket}						
\usepackage[utf8]{inputenc}			
\usepackage{color} 						
\usepackage{simplewick}
\usepackage{enumitem}

\newcommand{\eq}[1]{(\ref{#1})}
\newcommand{\Eq}[1]{Eq.~(\ref{#1})}
\newcommand{\Eqs}[1]{Eqs.~(\ref{#1})}
\newcommand{\Fig}[1]{Fig.~\ref{#1}}
\newcommand{\Sec}[1]{\S\ref{#1}}
\newcommand{\App}[1]{Appendix~\ref{#1}}

\newcommand{\pd}{\partial}
\newcommand{\del }{\vec{\nabla}}

\newcommand{\mc}[1]{\mathcal{#1}}

\newcommand{\msf}[1]{\mathsf{#1}}
\newcommand{\mcu}[1]{\mathscr{#1}}
\newcommand{\oper}[1]{\widehat{\mcu{#1}} }
\renewcommand{\vec}[1]{{\boldsymbol{#1}}}

\newcommand{\bralong}[1]{\bra{\, #1 \,}}
\newcommand{\ketlong}[1]{\ket{\, #1 \,}}
\newcommand{\inner}[1]{\braket{\, #1  \, }}
 
\newcommand{\moysin}[1]{	\boldsymbol{\{ } \! \{	 #1 \} \!	\boldsymbol{ \}} 	}
\newcommand{\moycos}[1]{ \boldsymbol{[} 	[	 #1 ] 	\boldsymbol{]}	}
\newcommand{\avg}[1]{ \boldsymbol{\langle} \! \langle   #1   \rangle\! \boldsymbol{\rangle} }

\newcommand{\fluctpsi}{\widetilde{\psi}}
\newcommand{\fluctw}{\widetilde{w}}
\newcommand{\fluctxi}{\widetilde{\xi}}
\newcommand{\ketx}{\ketlong{\msf{x}}}
\newcommand{\brax}{\bralong{\msf{x}}}
\newcommand{\kety}{\ketlong{\msf{y}}}
\newcommand{\bray}{\bralong{\msf{y}}}

\newcommand{\epnl}{\epsilon_{\rm nl}}
\newcommand{\ep}{\epsilon}

\newcommand{\ui}{{\rm i}}
\newcommand{\ue}{{\rm e}} 

\shorttitle{Wave kinetic eq. for inhomogeneous drift-wave turbulence}
\shortauthor{D.~E.~Ruiz, M.~E.~Glinsky, and I.~Y.~Dodin}

\title{Wave kinetic equation for inhomogeneous drift-wave turbulence \\ beyond the quasilinear approximation}

\author{
D.~E.~Ruiz\aff{1}	\corresp{\email{deruiz@sandia.gov}},
M.~E.~Glinsky\aff{1},
\and
I.~Y.~Dodin\aff{2,3}}

\affiliation{
\aff{1}Sandia National Laboratories, P.O. Box 5800, Albuquerque, New Mexico 87185, USA
\aff{2}Princeton Plasma Physics Laboratory, Princeton, New Jersey 08543, USA
\aff{3}Department of Astrophysical Sciences, Princeton University, Princeton, New Jersey 08544, USA
}

\begin{document}

\maketitle


\begin{abstract}
The formation of zonal flows from inhomogeneous drift-wave (DW) turbulence is often described using statistical theories derived within the quasilinear approximation. However, this approximation neglects wave--wave collisions. Hence, some important effects such as the Batchelor--Kraichnan inverse-energy cascade are not captured within this approach. Here we derive a wave kinetic equation that includes a DW collision operator in the presence of zonal flows. Our derivation makes use of the Weyl calculus, the quasinormal statistical closure, and the geometrical-optics approximation. The obtained model conserves both the total enstrophy and energy of the system. The derived DW collision operator breaks down at the Rayleigh--Kuo threshold. This threshold is missed by homogeneous-turbulence theory but expected from a full-wave quasilinear analysis. In the future, this theory might help better understand the interactions between drift waves and zonal flows, including the validity domain of the quasilinear approximation that is commonly used in literature.
\end{abstract}

\section{Introduction}
\label{sec:intro}

The interaction between drift-wave (DW) turbulence and zonal flows (ZFs) has been widely studied in plasma physics \citep{Diamond:2005br,Fujisawa:2009jc,Lin:1998je,Biglari:1990hx,Dorland:2000bb,Jenko:2000gn,Connaughton:2015kk}. In the context of magnetic fusion \citep{Fujisawa:2009jc,EUROfusionConsortium:2016bk,Conway:2005gq}, the spontaneous emergence of ZFs significantly affects the transport of energy, momentum, and particles. Understanding this phenomenon is critical to improving plasma confinement, but modeling the underlying physics is difficult. For example, direct numerical simulations of interacting DWs and ZFs strongly depend on the initial conditions and the external random forcing. Thus, statistical methods have been useful and are widely applied in the DW-turbulence research, even at the cost of introducing approximations. 

One particular statistical approach is the so-called quasilinear (QL) approximation \citep{Farrell:2003dm}, where the ZF equation is kept nonlinear and the equation for DWs is linearized. Among statistical QL theories, the \textit{wave kinetic equation} (WKE) is a popular model that captures the essential basic physics of DW turbulence, \eg the formation of ZFs \citep{Parker:2016eu,Ruiz:2016gv,Zhu:2018fd,Parker:2018eda,Zhu:2018hk, Zhu:2018gk,Diamond:2005br,Smolyakov:1999jk,Smolyakov:2000be,Malkov:2001kp,Malkov:2001hv,Diamond:1994fd,Kim:2003jf,Kaw:2002ku,Trines:2005in,Singh:2014bh}. The WKE has the intuitive form of the Liouville equation for the DW action density $J$ in the ray phase space \citep{Parker:2016eu,Ruiz:2016gv,Zhu:2018fd,Parker:2018eda,Zhu:2018hk, Zhu:2018gk}:
\begin{equation}
	\pd_t J +  \{J,\Omega\} = 2 \Gamma J ,
	\label{eq:intro:wke}
\end{equation}
where $\Omega$ is the local DW frequency, $\Gamma$ is a dissipation rate due to interactions with ZFs, and $\{\cdot,\cdot\}$ is the canonical Poisson bracket. (For the sake of clarity, terms related to external forcing and dissipation are omitted here.) However, as with all QL models, \Eq{eq:intro:wke} neglects nonlinear wave--wave scattering, and in consequence, is not able to capture the Batchelor--Kraichnan inverse-energy cascade \citep{Srinivasan:2012im} or produce the Kolmogorov--Zakharov spectra for DWs \citep{Connaughton:2015kk}. Hence, a question remains as to whether the existing WKE for inhomogeneous turbulence can be complemented with a wave--wave collision operator $C[J,J]$. The goal of this work is to calculate $C[J,J]$ explicitly. 

Starting from the generalized Hasegawa--Mima equations (gHME) \citep{Krommes:2000ec,Smolyakov:1999jk}, we derive a WKE with a DW collision operator $C[J,J]$ for inhomogeneous DW turbulence. Our derivation is based on the \emph{Weyl calculus} \citep{Weyl:1931uw}, which makes our approach similar to that in \citet{Ruiz:2016gv} where the QL approximation was used. However, in contrast to \citet{Ruiz:2016gv}, we do not rely on the QL approximation here but instead account for DW collisions perturbatively using the \textit{quasinormal approximation}. The main result of this work are \Eqs{eq:wke} and \eq{eq:wke_zonal}. In this final result, DWs are modeled in a similar manner as in \Eq{eq:intro:wke}. The difference is that \Eq{eq:wke} includes nonlinear DW scattering, which is described by a wave--wave collision operator $C[J,J]$ that is bilinear in the DW wave-action density $J$. The resulting model conserves the two nonlinear invariants of the gHME, which are the total enstrophy and the total energy. 

The present formulation is fundamentally different from the previously reported homogeneous weak-wave-turbulence models for DW turbulence \citep{Connaughton:2015kk,Krommes:2002hva}. While DWs are described as an incoherent fluctuating field as usual, ZFs are now treated as coherent structures, which are missed in homogeneous-turbulence theory. The obtained model motivates future investigations of the effects of nonlinear wave--wave scattering on DW--ZF interactions, in particular, the spontaneous emergence of ZFs and the eventual saturation of the ZFs and the DW spectra. This theory might also help better understand the validity domain of the quasilinear approach to DW turbulence that has been commonly used in the literature.

The present work is organized as follows. In \Sec{sec:basic}, we introduce the gHME and obtain the governing equations for the mean and fluctuating components of the fields.  In \Sec{sec:stats}, we introduce the quasinormal statistical closure to obtain a closed equation for the correlation operator describing the vorticity fluctuations. In \Sec{sec:WKE}, we project the equations into the DW-ray phase space and obtain the collisional WKE. In \Sec{sec:discussion}, we discuss the obtained equations, their conservation properties, and their relation to previous models. Final conclusions and remarks are given in \Sec{sec:conclusions}. In \App{app:conventions}, we give a brief introduction to the Weyl calculus and also define the zonal average of an arbitrary operator. In \App{app:auxiliary}, we present some auxiliary calculations.

\section{Basic equations}
\label{sec:basic}

\subsection{The generalized Hasegawa--Mima model}
\label{sec:gHME}

We consider a magnetized plasma in a uniform magnetic field in the $z$ direction and with an equilibrium local gradient of the plasma density. Upon assuming a quasi-adiabatic response for the electrons and a fluid description for the ion dynamics, we model electrostatic two-dimensional (2-D) turbulent flows in the $xy$ plane using the gHME \citep{Krommes:2000ec,Smolyakov:1999jk}, which in normalized units is given by
\begin{gather}
	\pd_t w + \vec{v}\cdot \del w + \beta\, \pd_x \psi = Q.
	\label{eq:basic_hm}
\end{gather}
Here $\vec{x} = (x, y)$ is a 2-D coordinate, the $x$ axis is the ZF direction, and the $y$ axis is the direction of the local gradient of the background plasma density, which is measured by the constant $\beta$. In \Eq{eq:basic_hm}, time is measured in units of the inverse ion cyclotron frequency while length is measured in units of the ion sound radius. [See, \eg \citet{Zhu:2018fd} for more details.]
The function $\psi(t,\vec{x})$ is the electric potential, $\vec{v} \doteq \vec{e}_z \times \del \psi$ is the ion fluid velocity on the $xy$ plane, and $\vec{e}_z$ is a unit vector normal to this plane. (The symbol $\doteq$ denotes definitions.) The generalized vorticity $w(t,\vec{x})$ is given by
\begin{equation}
	w(t,\vec{x}) \doteq (\nabla^2 - L_{\rm D}^{-2} \widehat{a})\psi,
\end{equation}
where $\widehat{a}$ is an operator such that $\widehat{a} = 1$ in parts of the spectrum corresponding to DWs and $\widehat{a} = 0$ in those corresponding to ZFs. The constant $L_{\rm D}$ is the ion sound radius. ($L_{\rm D} = 1$ in normalized units.) The term $Q(t,\vec{x})$ in \Eq{eq:basic_hm} represents external forces and dissipation. Notably, \Eq{eq:basic_hm} with $L_{\rm D} \to \infty$ also describes Rossby turbulence in planetary atmospheres \citep{Farrell:2003dm, Farrell:2007fq, Marston:2008gx, Srinivasan:2012im, AitChaalal:2016jx}.

For isolated systems, where $Q=0$, \Eq{eq:basic_hm} conserves two nonlinear invariants: the enstrophy $\mc{Z}$ and the energy $\mc{E}$ (strictly speaking, free energy). These are defined as
\begin{gather}\label{eq:ZE}
	\mc{Z}(t) \doteq \frac{1}{2}\int \mathrm{d}^2\vec{x} \,	w^2	 ,
	\quad 
	\mc{E}(t) \doteq - \frac{1}{2} \int \mathrm{d}^2\vec{x} \, w \psi .
\end{gather}
The statistical model that we shall derive conserves both of these nonlinear invariants.

\subsection{Separating the mean and fluctuating components of the fields}
\label{sec:mean_fluct}

Let us decompose the fields $\psi$ and $w$ into their mean and fluctuating components, which will be denoted by bars and tildes, respectively. For any arbitrary field $g(t,\vec{x})$, the mean part is defined as $\overline{g}(t,y) \doteq  \int \mathrm{d}x \, \avg{g}/L_x$, where $L_x$ is the system's length along $x$ and $\avg{ \cdot }$ denotes the statistical average over realizations of initial conditions or of the external random forcing. The mean part will describe the coherent ZF dynamics. In contrast, the fluctuating quantities will describe the incoherent DW dynamics.

We consider the fluctuating quantities to be small in amplitude. The small-amplitude ordering is denoted by using the small dimensionless parameter $\epnl \ll 1$. In particular,
\begin{equation}
	w = \overline{w} + \epnl  \,   \fluctw,
	\quad \quad
	\vec{v} = U \vec{e}_x + \epnl \widetilde{\vec{v}}	,
\end{equation}
where $U(t,y) \doteq - \pd_y \overline{\psi}$ is the ZF velocity field and the two components of the generalized vorticity are related to $\psi$ as \citep{Parker:2013hy}

\begin{gather}
	\overline{w} = \nabla^2 \overline{\psi}, 
	\quad \quad
	\fluctw = \nabla_{\rm D}^2 \widetilde{\psi}.
\end{gather}
Here $\nabla_{\rm D}^2 \doteq \nabla^2 - 1$. Also, $\overline{\fluctw}=0$ and $\overline{\widetilde{\vec{v}}} = 0$ by definition.

From \Eq{eq:basic_hm}, we then derive the governing equations for the fluctuating and the mean fields \citep{Srinivasan:2012im}:
\begin{subequations} \label{eq:basic_mean_fluctuations}
	\begin{gather}
		\pd_t \fluctw 
				+ U \pd_x \fluctw 
				+ [\beta - (\pd_y^2 U)] \pd_x \widetilde{\psi} 
				+ \epnl \, f_{\rm eddy}
				=  \widetilde{Q} , \\
		\pd_t U 
				+ \mu_{\rm zf} U 
				+ \epnl^2\,  \pd_y \overline{\widetilde{v}_x \widetilde{v}_y} = \overline{Q},
	\end{gather}
\end{subequations}
where $\smash{f_{\rm eddy}(t,\vec{x}) \doteq \widetilde{\vec{v}} \cdot \del \fluctw - \overline{\widetilde{\vec{v}} \cdot \del \fluctw}}$ is a nonlinear term representing eddy--eddy interactions and is responsible for the Batchelor--Kraichnan inverse-energy cascade \citep{Srinivasan:2012im}. As seen from \Eqs{eq:basic_mean_fluctuations}, the parameter $\epnl$ denotes the smallness of the nonlinear coupling terms. Also note that we considered the ordering $\smash{Q=\overline{Q} + \epnl \, \widetilde{Q}}$.

\subsection{Temporal and spatial scale separation of fluctuating and mean fields}
\label{sec:GO}

In the model derived below, we shall assume that there is a temporal and spatial scale separation between the DW and ZF fields. Specifically, let $\tau_{\rm dw}$ and $\lambda_{\rm dw}$ respectively denote the characteristic period and wavelength of the DWs. In a similar manner, the characteristic time and length scales of the ZFs are given by $T_{\rm zf}$ and $L_{\rm zf}$, respectively. Upon following the discussion in \citet{Ruiz:2016gv}, we shall characterize the scale separation between DWs and ZFs by introducing the geometrical-optics (GO) parameter
\begin{gather}
 		\ep \doteq 
 			\mathrm{max}	\left(
 				\frac{\tau_{\rm dw}}{T_{\rm zf}},
 				\frac{\lambda_{\rm dw}}{L_{\rm zf}},
	 			\frac{L_{\rm D}}{L_{\rm zf}}
 				\right)  \ll 1.
  		\label{eq:GO_condition}
\end{gather}

In addition, we assume that the DWs are weakly damped and weakly dissipated so that $\smash{\widetilde{Q} = \ep\, \widetilde{\xi} - \ep\,  \mu_{\rm dw} \fluctw}$, where $\widetilde{\xi}$ is some white-noise external forcing with zero mean and $\mu_{\rm dw}$ is intended to emulate the dissipation of DWs caused by the external environment. The external dissipation and random forcing are scaled using the GO parameter \citep{McDonald:1985ib}. For the mean quantities, we consider $\smash{\overline{Q} = - \mu_{\rm zf} \, \overline{w}}$, where $\mu_{\rm zf}$ represents dissipation on the mean flows.

Upon using the orderings above, we write \Eqs{eq:basic_mean_fluctuations} as
\begin{subequations}	\label{eq:basic_main}
	\begin{gather}
		\ep \pd_t \fluctw 
				+ \ep U \pd_x \fluctw 
				+ \ep [\beta - (\pd_y^2 U)] \pd_x \widetilde{\psi} 
				+ \ep \mu_{\rm dw} \fluctw 
				+  \epnl f_{\rm eddy}
				=  \ep \widetilde{\xi} ,
				\label{eq:basic_w} \\
		\pd_t U 
				+ \mu_{\rm zf} U 
				+ \epnl^2 \, \pd_y \overline{\widetilde{v}_x \widetilde{v}_y} = 0, 
				\label{eq:basic_U}
	\end{gather}
\end{subequations}
where the GO parameter $\ep$ was inserted in front of the partial derivatives acting on fluctuating quantities in order to denote the small-scale ordering. (Here $\ep$ serves the same role as $\hbar$ in quantum mechanics.) Note that $ \smash{ \widetilde{\vec{v}} = \vec{e}_z \times \ep \del \fluctpsi } $, $\smash{ \fluctw = ( \ep^2 \nabla^2 - 1) \fluctpsi }$, and $f_{\rm eddy} = \smash{ \widetilde{\vec{v}} \cdot \ep \del \fluctw -  \overline{\widetilde{\vec{v}} \cdot \ep \del \fluctw}}$. Also, note $\smash{ \pd_y \overline{\widetilde{v}_x \widetilde{v}_y} } \sim O(\ep^0)$ since $\smash{ \overline{\widetilde{v}_x \widetilde{v}_y} }$ is a mean quantity. Equations \eq{eq:basic_main} govern the dynamics of the fluctuating vorticity $\fluctw(t,\vec{x})$ and of the ZF velocity $U(t,y)$. Although we have presented our ordering assumptions in \Sec{sec:mean_fluct} and \Sec{sec:GO}, no approximations have been adopted yet, so these equations are equivalent to the gHME.

\subsection{Abstract vector representation}

Let us write the fluctuating fields as elements of an abstract Hilbert space $L^2(\mathbb{R}^3)$ of wave states with inner product \citep{Dodin:2014hw,Littlejohn:1993bd}
\begin{equation}
	\inner{ \phi \mid \psi } = \int \mathrm{d} t \, \mathrm{d}^2 \vec{x} \, \phi^*(t,\vec{x}) \, \psi(t,\vec{x}),
	\label{eq:abstract_inner}
\end{equation}
where the integrals are taken over $\mathbb{R}^3$. Let $\ketlong{t,\vec{x}}$ be the eigenstates of the time and position operators $\widehat{t}$ and $\widehat{\vec{x}}$. [Considering $\widehat{t}$ as an operator will allow us to write the statistical closure in \Sec{sec:stats} in a simple manner.] Hence, $\fluctw(t,\vec{x})$ is written as $\fluctw(t,\vec{x})=\inner{t,\vec{x} \mid \fluctw}$. Since $\fluctw(t,\vec{x})$ is real, then $\inner{\fluctw \mid t,\vec{x}} = \inner{ t,\vec{x} \mid \fluctw}$. In the following, we shall use the notation $\msf{x} \doteq(t,\vec{x})$ to denote spacetime coordinates. Likewise, $\ketlong{ \msf{x} } \doteq \ketlong{t,\vec{x}}$ is the corresponding space-time eigenstate, and $\mathrm{d}^3 \msf{x} \doteq \mathrm{d}t \, \mathrm{d}^2 \vec{x}$ is the spacetime differential volume.

In addition to the time and position operators, we introduce the frequency operator $\widehat{\omega}$, such that $\widehat{\omega} \doteq \ui \ep \pd_t$ in the coordinate representation. Likewise, the wavevector operator $\oper{\vec{k}}$ is defined as $\oper{\vec{k}} \doteq - \ui \ep \del$. In particular, these operators are Hermitian, and one has $\inner{\msf{x} \mid \widehat{\omega} \mid \fluctw} = \ui \ep \pd_t \fluctw$ and $\inner{ \msf{x} \mid \widehat{\vec{k}} \mid \fluctw} = -\ui \ep \del \fluctw$. Using the relation between the fluctuating generalized vorticity and electric potential, one has $\ket{\fluctw} = - \widehat{k}_{\rm D}^2 \ket{\widetilde{\psi}}$, where
\begin{gather}
 	\widehat{k}_{\rm D}^2 \doteq \widehat{k}^2 + 1, 
 	\quad \quad
 	\widehat{k}^2 \doteq \oper{\vec{k}} \cdot \oper{\vec{k}}.
 	\label{eq:pbar}
\end{gather}

Let us write the fluctuating quantities appearing in \Eq{eq:basic_main} in terms of the abstract states and operators. One finds that \Eq{eq:basic_w} can be written as
\begin{equation}
	\oper{D} \ketlong{\fluctw} 
			= 		\frac{\ui \epnl}{2} \ketlong{ f_{\rm nl}[\fluctw, \fluctw ]}
				- 	\frac{\ui \epnl}{2} \ketlong{ \overline{f_{\rm nl}[\fluctw, \fluctw ]}} 
				+ \ui \ep \ketlong{\widetilde{\xi}}.
	\label{eq:abstract_fluct}
\end{equation}
Here $\oper{D}$ is the DW dispersion operator
\begin{equation}
	\oper{D} 	\doteq		
							\widehat{\omega} 
							- \widehat{U} \widehat{k}_x 
							+ (\beta - \widehat{U}'') \widehat{k}_x \widehat{k}_{\rm D}^{-2} 
							+ \ui \ep \mu_{\rm dw},
\end{equation}
where $\smash{\widehat{U} \doteq U(\widehat{t},\widehat{y})}$, and the prime above $U$ henceforth denotes $\pd_y$; in particular, $\smash{\widehat{U}'' \doteq \pd_y^2 \, U( \widehat{t} ,\widehat{y})}$. Also, $\smash{\ketlong{\widetilde{\xi}}}$ is the ket corresponding to the random forcing $\smash{\widetilde{\xi}}$. It is to be noted that the $\smash{\oper{D}}$ includes the nonlinear coupling between the ZFs and the DWs.

Additionally in \Eq{eq:abstract_fluct}, the kets $\ketlong{ f_{\rm nl}[\phi,\psi]}$ and $\ketlong{ \overline{f_{\rm nl}[\phi,\psi ]}}$ are given by
\begin{align}
	\ketlong{ f_{\rm nl}[\phi,\psi]} 
			&	\doteq \int 	\mathrm{d}^3 \msf{x}  \, 	\ketx \inner{\phi \mid \oper{K} (\msf{x}) \mid \psi } ,	
			\label{eq:abstract_f_nl} \\
	\ketlong{ \overline{f_{\rm nl}[\phi,\psi ]} } 
			&	\doteq \int 	\mathrm{d}^3 \msf{x}  \, 	\ketx \overline{  \inner{\phi \mid \oper{K} (\msf{x}) \mid \psi }} ,
\end{align}
where $\oper{K} (\msf{x})$ is an operator describing the nonlinear eddy--eddy interactions
\begin{equation}
	\oper{K}(\msf{x}) 	\doteq	
						\oper{L}_j \ketx \brax \oper{R}_j 
						+ \oper{R}_j \ketx \brax \oper{L}_j	.
	\label{eq:abstract_K}
\end{equation}
(The summation over repeating indices is assumed.) The operators $\smash{\oper{L}_j}$ and $\smash{\oper{R}_j}$ are given by
\begin{equation}	\label{eq:abstract_oper}
	\oper{L}_j \doteq (\vec{e}_z \times \widehat{\vec{k}})_j k_{\rm D}^{-2}, \quad \quad \quad 
	\oper{R}_j \doteq \widehat{\vec{k}}_j.
\end{equation}
Indeed, we can verify that $\ketlong{ f_{\rm nl}[\fluctw,\fluctw]}$ represents nonlinear eddy-eddy interactions. When projecting onto the coordinate eigenstates, one obtains
\begin{align}
	 \inner{ \msf{x}  \mid f_{\rm nl}[\fluctw,\fluctw] } 
		&	=	\inner{\fluctw \mid \oper{K} (\msf{x}) \mid \fluctw }	\notag \\
		&	=	\inner{\fluctw \mid \widehat{k}_{\rm D}^{-2}(\vec{e}_z \times \widehat{\vec{k}})_j \mid \msf{x} }  
				\inner{ \msf{x} \mid \widehat{\vec{k}}_j \mid \fluctw }	
				+ \inner{\fluctw  \mid \widehat{\vec{k}}_j  \mid \msf{x} } 
				\inner{\msf{x}  \mid  (\vec{e}_z \times \widehat{\vec{k}})_j \widetilde{k}_{\rm D}^{-2}  \mid \fluctw }
				\notag \\
		&	=	- ( \inner{ \msf{x}  \mid (\vec{e}_z \times \widehat{\vec{k}})_j \mid \widetilde{\psi} }   )^*
				\inner{ \msf{x} \mid \widehat{\vec{k}}_j \mid \fluctw }
				- ( \inner{ \msf{x}  \mid \widehat{\vec{k}}_j  \mid \fluctw }   )^*
				\inner{ \msf{x} \mid (\vec{e}_z \times \widehat{\vec{k}})_j  \mid \fluctpsi }
				\notag \\
		&	=	-2 (\vec{e}_z \times \ep \del \fluctpsi) \cdot \ep \del \fluctw \notag \\
		&	=	-2 \widetilde{\vec{v}} \times \ep \del \fluctw,
\end{align}
which is one of the nonlinear terms appearing in \Eq{eq:basic_w}. Here we substituted the coordinate representation for the wavevector operator and used $\smash{\ket{\fluctw} = - \widehat{k}_{\rm D}^2 \ket{\fluctpsi}}$. Note that the factor 2 appears because we have written the ket $\ketlong{ f_{\rm nl}[\phi,\psi]}$ in a symmetric form so that
\begin{equation}
	\ketlong{ f_{\rm nl}[\phi,\psi]}  = \ketlong{ f_{\rm nl}[\psi,\phi]}
	\label{eq:basic_symmetry}
\end{equation}
for any two real fields $\phi$ and $\psi$.

Following \citet{Ruiz:2016gv}, the fluctuating terms appearing in \Eq{eq:basic_U} can also be rewritten in the abstract representation. Upon noting that $\widetilde{v}_x \widetilde{v}_y = (-\ep \pd_y \fluctpsi) ( \ep \pd_x \fluctpsi ) = -\inner{ \msf{ x} \mid \widehat{k}_x \mid \fluctpsi } \inner{ \fluctpsi \mid \widehat{k}_y \mid \msf{x} }$, one obtains
\begin{equation}
	\pd_t U + \mu_{\rm zf} U = \epnl^2 \, \pd_y \overline{
													\inner{\msf{x} \mid \widehat{k}_x \widehat{k}_{\rm D}^{-2} \mid \fluctw} 
													\inner{\fluctw \mid \widehat{k}_{\rm D}^{-2} \widehat{k}_y \mid \msf{x} } }.
	\label{eq:Dirac_mean}
\end{equation}

\section{Statistical closure}
\label{sec:stats}

\subsection{Statistical-closure problem}

Let us now introduce the correlation operator for the fluctuating vorticity field:
\begin{equation}
	\oper{W} \doteq \overline{ \ketlong{ \fluctw} \bralong{ \fluctw }},
	\label{eq:stats_Wigner}
\end{equation}
where the Hermitian operator $\ketlong{\fluctw} \bralong{\fluctw}$ can be interpreted as the \emph{fluctuating-vorticity density operator} by analogy with quantum mechanics \citep{Ruiz:2016gv}. It is to be noted that $\smash{\oper{W}}$ is defined as the zonal average of the fluctuating-vorticity density operator. The zonal average of an abstract operator is discussed in \App{app:zonal_average}. By using the identity \eq{eq:zonal_avg_identity}, one can write \Eq{eq:Dirac_mean} in terms of the correlation operator:
\begin{equation}
	\pd_t U + \mu_{\rm zf} U = \epnl^2 \, \pd_y
													\inner{\msf{x} \mid 
														\widehat{k}_x \widehat{k}_{\rm D}^{-2} 
														\oper{W}
														\widehat{k}_{\rm D}^{-2} \widehat{k}_y 
													\mid \msf{x} } .
	\label{eq:stats_mean}
\end{equation}
%

Now, let us obtain the governing equation for $\oper{W}$. Multiplying \Eq{eq:abstract_fluct} by $\bralong{\fluctw}$ from the right and zonal averaging leads to an equation for the correlation operator:
\begin{align}
	\oper{D}  \oper{W}
				=	& \, 	\frac{\ui \epnl}{2}	\,
							\overline{ \ketlong{ f_{\rm nl} [\fluctw ,\fluctw ] } \bralong{\fluctw}    } 
					 +\ui \ep \, \overline{ \ketlong{\widetilde{\xi}} \bralong{\fluctw} }.
	\label{eq:stats_eq_original}
\end{align}
Subtracting from \Eq{eq:stats_eq_original} its Hermitian conjugate gives
\begin{gather}
		[\oper{D}_{\rm H}, \oper{W} ]_- 
		+ \ui [\oper{D}_{\rm A}, \oper{W} ]_+ 
			 	=	\ui \epnl \,
						\big[ 	\overline{   \ketlong{ f_{\rm nl} [\fluctw ,\fluctw ] }   \bralong{\fluctw}   }
						\big]_{\rm H}				
					+ 2 \ui \ep \,
						\big[ \overline{ \ketlong{\widetilde{\xi}} \bralong{\fluctw}  } \big]_{\rm H},
		\label{eq:stats_main}
\end{gather}
where the subscripts ``H" and ``A" denote the Hermitian and anti-Hermitian parts of an arbitrary operator; i.e., $\oper{A}_{\rm H} \doteq (\oper{A} + \oper{A}^\dag)/2$ and $\oper{A}_{\rm A} \doteq  (\oper{A} - \oper{A}^\dag)/(2\ui)$. Specifically, the operator $\oper{D}$ is decomposed as $\oper{D}  = \oper{D}_{\rm H} + \ui \oper{D}_{\rm A}$, where
\begin{subequations}	\label{eq:D_H_A}
\begin{align}
	\oper{D}_{\rm H} 
				&		=			\widehat{\omega} - \widehat{U} \widehat{k}_x 
								+[ \beta - \widehat{U}'', \widehat{k}_x  \widehat{k}_{\rm D}^{-2} ]_+/2,	
				\label{eq:D_H}					\\
	\oper{D}_{\rm A} 
				&		=		\ui	[ \widehat{U}'', \widehat{k}_x  \widehat{k}_{\rm D}^{-2} ]_-/2 
								+ \ep \mu_{\rm dw} .
				\label{eq:D_A}
\end{align}
\end{subequations}
Both $\oper{D}_{\rm H}$ and $\oper{D}_{\rm A}$ are Hermitian. Also, $[ \cdot , \cdot ]_{\mp}$ denote the commutators and anticommutators; i.e., $\smash{[\oper{A}, \oper{B}]_- = \oper{A} \oper{B} - \oper{B} \oper{A}}$ and $\smash{[\oper{A}, \oper{B}]_+ = \oper{A} \oper{B} + \oper{B} \oper{A}}$. Aside from the additional first term on the right-hand side of \Eq{eq:stats_main}, this result is similar to that obtained in \citet{Ruiz:2016gv}.

Equations \eq{eq:stats_mean} and \eq{eq:stats_main} are not closed. The left-hand side of \Eq{eq:stats_main} is written in terms of $\smash{\oper{W}}$, which is bilinear in the fluctuating vorticity. However, the right-hand side of \Eq{eq:stats_main} contains terms that are linear and cubic with respect to $\fluctw$. This is the fundamental \emph{statistical closure problem} \citep{frisch1995turbulence,kraichnan2013closure,Krommes:2002hva}. The next step is to introduce a statistical closure in order to express \Eq{eq:stats_main} in terms of the correlation operator $\smash{\oper{W}}$ only.

\subsection{A comment on the quasilinear approximation}
\label{sec:QL}

One possibility is to neglect the first term on the right-hand side of \Eq{eq:stats_main}. Then, one linearizes \Eq{eq:abstract_fluct} so that
\begin{equation}
	\oper{D} \ketlong{ \fluctw } \simeq \ui \ep \ketlong{ \fluctxi }.
	\label{eq:stats_QL_aux}
\end{equation}
After formally inverting $\oper{D}$, we have $\ketlong{ \fluctw } \simeq \ui \ep \oper{D}^{-1} \ketlong{ \fluctxi } $. Substituting into \Eq{eq:stats_main} leads to the closed equation
\begin{equation}
	[\oper{D}_{\rm H}, \oper{W} ]_- + \ui [\oper{D}_{\rm A}, \oper{W} ]_+ 
		=	2 \ui \ep^2  \,
			\big[ \oper{S} (\oper{D}^{-1})^\dag \big]_{\rm A},
	\label{eq:stats_quasilinear}
\end{equation}
where $\oper{S}	\doteq \overline{  \ketlong{ \fluctxi } \bralong{ \fluctxi } }$ is the zonal-averaged density operator associated with the random external forcing. We also used the fact that, for any operator $\oper{A}$, one has $(- \ui \, \oper{A})_{\rm H}=\oper{A}_{\rm A}$.

Equations \eq{eq:stats_mean} and \eq{eq:stats_quasilinear} now form a closed system. In this approximation, the equation for the DW fluctuations was linearized in \Eq{eq:stats_QL_aux}, but the DW nonlinearity is only kept in the equation for the mean field [\Eq{eq:stats_mean}]. This constitutes the QL approximation. If one projects \Eq{eq:stats_quasilinear} on the double-physical coordinate space $(\msf{x}, \msf{x'})$ using multiplication by $\bralong{t,\vec{x}}$ and $\ketlong{t,\vec{x}'}$, then one obtains the so-called CE2 equations \citep{Farrell:2003dm, Farrell:2007fq, Marston:2008gx, Srinivasan:2012im, AitChaalal:2016jx}. Likewise, if one projects \Eq{eq:stats_quasilinear} on the ray space $(t,\vec{x},\omega,\vec{k})$ using the Weyl transform \citep{Weyl:1931uw}, then one obtains the Wigner--Moyal model discussed in \citet{Ruiz:2016gv}, \citet{Parker:2018eda}, and \citet{Zhu:2018fd}.

\subsection{A statistical closure beyond the quasilinear approximation}
\label{sec:closure}

We extend our theory beyond the QL approximation in order to retain wave--wave scattering, namely, the term $f_{\rm nl}$ in \Eq{eq:stats_main}. Let us separate $\fluctw$ into three components:

\begin{equation}
	\ketlong{\fluctw} = \ketlong{\fluctw_0} + \epnl \ketlong{ \widetilde{\phi} } + \ep \ketlong{\widetilde{\varphi}}.
	\label{eq:closure_separation}
\end{equation}
Here $\ketlong{\fluctw_0} = O(1)$ is chosen to satisfy the linear part of \Eq{eq:abstract_fluct}; i.e., $\oper{D} \ketlong{\fluctw_0}  =0$. The fluctuations in $\fluctw_0$ are due to random initial conditions, whose statistics are considered to be uncorrelated to those of the random forcing $\fluctxi$. When formally inverting $\oper{D}$ in \Eq{eq:abstract_fluct}, one finds that, to lowest order in $\epnl$ and $\ep$,
\begin{subequations}	\label{eq:closure_aux_var}
	\begin{gather}
			\ketlong{ \widetilde{\phi} }  
					\simeq	\frac{\ui }{2} \,
								\oper{D}^{-1} 
							\big\{   \ketlong{ f_{\rm nl} [\fluctw_0 ,\fluctw_0 ] } 
					 					-	\overline{  \ketlong{ f_{\rm nl} [\fluctw_0 ,\fluctw_0 ] } } \big\} ,		\label{eq:closure_phi}		\\
			\ketlong{\widetilde{\varphi}}
					\simeq	i \oper{D}^{-1} \ketlong{ \fluctxi }  .
		\label{eq:closure_varphi}		
	\end{gather}
\end{subequations}
Substituting \Eqs{eq:closure_separation} and \eq{eq:closure_aux_var} into \Eq{eq:stats_main} leads to
\begin{align}
	[\oper{D}_{\rm H}, \oper{W} ]_- 	+ \ui [\oper{D}_{\rm A}, \oper{W} ]_+ 
		=	&		\, \ui \epnl  \big\{ 	\overline{
								\ketlong{ f_{\rm nl} [\fluctw_0 ,\fluctw_0 ] }  
								\bralong{\fluctw_0}	}	\big\}_{\rm H}	
					+ 2\ui  \epnl^2  \big\{ 	\overline{
								\ketlong{ f_{\rm nl} [\widetilde{\phi} ,\fluctw_0 ] }  
								\bralong{\fluctw_0}	}	\big\}_{\rm H}	\notag \\
			&	+ \ui \epnl^2		\big\{	\overline{   
								 \ketlong{ f_{\rm nl} [\fluctw_0 ,\fluctw_0 ] }  
								 \bralong{\widetilde{\phi}} } \big\}_{\rm H}	
					+ 2 \ui \ep^2   \big[  \oper{S}  (\oper{D}^{-1})^\dag \big]_{\rm A} \notag \\
			&		+ O(\epnl^3, \epnl^2 \ep, \epnl \ep^2).
		\label{eq:closure_main_aux}
\end{align}
Here we neglected $O(\epnl^3, \epnl^2 \ep, \epnl \ep^2)$ terms, \eg $\epnl^3 \ketlong{ f_{\rm nl} [\fluctw_0 ,\widetilde{\phi} ] }  \bralong{\widetilde{\phi}} $ and $\epnl^2 \ep \ketlong{ f_{\rm nl} [\fluctw_0 , \widetilde{\phi} ] }  \bralong{\widetilde{\varphi}} $. Note that the zonal averages of quantities involving both $\fluctw_0$ and $\widetilde{ \xi }$ (\eg $\smash{\epnl  \ketlong{\fluctxi }\bralong{\fluctw_0 } }$ and $\smash{ \epnl \ep  \ketlong{ f_{\rm nl} [\fluctw_0 , \fluctw_0 ] }  \bralong{\fluctxi}  }$) are zero since the statistics of $\fluctw_0$ and $\fluctxi$ are independent. The factor of two in the second term on the right-hand side of \Eq{eq:closure_main_aux} is due to the symmetry property \eq{eq:basic_symmetry}.

Now, let us explicitly calculate the statistical average of the nonlinear terms in \Eq{eq:closure_main_aux}. To do this, we shall use the quasinormal approximation which expresses higher-order $(n\geq 3) $ statistical moments of $\fluctw_0$ in terms of the lower-order moments. A further discussion on the validity of this approximation and its relation to the widely used random-phase approximation in homogeneous wave turbulence theory is given in \S\ref{sec:discussion}. When adopting the quasinormal approximation, one specifically obtains
\begin{equation}
	\overline{ \fluctw_0(\msf{x}_1)  \fluctw_0(\msf{x}_2) \fluctw_0(\msf{x}_3)  } =0,
	\label{eq:closure_quasi_cubic}
\end{equation}
\begin{align}
	 \overline{ \fluctw_0(\msf{x}_1 ) \fluctw_0(\msf{x}_2) \fluctw_0(\msf{x}_3) \fluctw_0(\msf{x}_4) } 
	& 				=			\inner{ \msf{x}_1 \mid \oper{W}_0 \mid \msf{x}_2 } 
							\, \inner{ \msf{x}_3 \mid \oper{W}_0 \mid \msf{x}_4 }   \notag \\
	&		\quad 		
				+			\inner{ \msf{x}_1 \mid \oper{W}_0 \mid \msf{x}_3 } 
							\, \inner{ \msf{x}_2 \mid \oper{W}_0 \mid \msf{x}_4 }  		
                            \notag \\
    &		\quad                      
                +			\inner{ \msf{x}_1 \mid \oper{W}_0 \mid \msf{x}_4 } 
							\, \inner{ \msf{x}_2 \mid \oper{W}_0 \mid \msf{x}_3 }  ,
	\label{eq:closure_quasi_quartic}
\end{align}
where $\oper{W}_0 \doteq \overline{ \ketlong{\fluctw_0}  \bralong{\fluctw_0} }$ and $\msf{x}_i=(t_i, \vec{x}_i)$ denotes a spacetime coordinate. 

Upon substituting \Eq{eq:closure_quasi_cubic}, one finds that the first term appearing in the right-hand side of \Eq{eq:closure_main_aux} vanishes; i.e., $\overline{\ketlong{ f_{\rm nl} [\fluctw_0 ,\fluctw_0 ] }  \bralong{\fluctw_0}	}=0$. After substituting \Eqs{eq:abstract_f_nl} and \eq{eq:closure_phi}, one obtains the following expression for the second term on the right-hand side of \Eq{eq:closure_main_aux}:
\begin{multline}
	\overline{  \ketlong{ f_{\rm nl} [\widetilde{\phi} ,\fluctw_0 ] }  
	 		\bralong{\fluctw_0} 	}
			=		\int \mathrm{d}^3 \msf{x} \, 
					\ketx	\overline{     \inner{ \widetilde{\phi}  \mid \oper{K}(\msf{x}) \mid \fluctw_0}  \bralong{\fluctw_0}	}	
			=		-\frac{\ui }{2}
							\int 	\mathrm{d}^3 \msf{x} \,  \mathrm{d}^3 \msf{y} \, 	
							\ketx 
				\\ 	\times
							\overline{     
								\big[  \inner{ \fluctw_0 \mid \oper{K}^\dag (\msf{y}) \mid \fluctw_0 } 
					 					-	\overline{ \inner{ \fluctw_0 \mid \oper{K}^\dag(\msf{y}) \mid \fluctw_0} } \big]		
								\inner{ \msf{y} \mid (\oper{D}^{-1})^\dag \,	 \oper{K}(\msf{x}) \mid \fluctw_0}  \bralong{\fluctw_0}	}	,
	\label{eq:closure_first}	
\end{multline}
where $\ketlong{\msf{x}}$ and $\ketlong{\msf{y}}$ are eigenstates of the coordinate operator. Inserting \Eq{eq:closure_quasi_quartic} in the abstract representation gives
\begin{align}
	& 	\overline{   \, \,  \big[  \inner{ \fluctw_0 \mid \oper{K}^\dag (\msf{y}) \mid \fluctw_0} 
					 					-	\overline{ \inner{ \fluctw_0 \mid \oper{K}^\dag(\msf{y}) \mid \fluctw_0} } \big]		
								\inner{ \msf{y} \mid (\oper{D}^{-1})^\dag \,	 \oper{K}(\msf{x}) \mid \fluctw_0}  \bralong{\fluctw_0}	\, \, }	
				\notag \\
		&			\quad \quad
			=	\contraction{\langle \, \,}{\fluctw_0}{\mid \oper{K}^\dag(\msf{y}) \mid \fluctw_0 \rangle \langle  \rangle \, \msf{y} \mid (\oper{D}^{-1})^\dag \,	 \oper{K}(\msf{x}) \mid  }{ \fluctw_0}
				\contraction[2ex]{\langle \, \fluctw_0 \mid \oper{K}^\dag(\msf{y}) \mid }{\fluctw_0 \,\,}{ \, \rangle \,\,\, \langle \,  \mid (\oper{D}^{-1})^\dag \,	 \oper{K}(\msf{x}) \mid \fluctw_0 \, \rangle \, \, \, \langle \, }{ \fluctw_0}
				\inner{ \fluctw_0 \mid \oper{K}^\dag(\msf{y}) \mid \fluctw_0}
				\inner{ \msf{y} \mid (\oper{D}^{-1})^\dag \,	 \oper{K}(\msf{x}) \mid \fluctw_0}  \bralong{\fluctw_0}	\notag \\ 
		&		\quad \quad \quad +
				\contraction{\langle \, \fluctw_0 \mid \oper{K}^\dag(\msf{y}) \mid \,}{\fluctw_0 }{ \, \rangle \,\,\,\, \langle \,  \mid (\oper{D}^{-1})^\dag \,	 \oper{K}(\msf{x}) \mid \,\, }{ \fluctw_0}
				\contraction[2ex]{\langle \, \,}{\fluctw_0}{\mid \oper{K}^\dag(\msf{y}) \mid \fluctw_0 \rangle \langle \, \rangle \, \msf{y} \mid (\oper{D}^{-1})^\dag \,	 \oper{K}(\msf{x}) \mid  \fluctw_0 \, \rangle  \langle \, }{ \fluctw_0}				
				\inner{ \fluctw_0 \mid \oper{K}^\dag(\msf{y}) \mid \fluctw_0}
				\inner{ \msf{y} \mid (\oper{D}^{-1})^\dag \,	 \oper{K}(\msf{x}) \mid \fluctw_0}  \bralong{\fluctw_0}	
				\notag \\
		&			\quad \quad
			= 2 
				\contraction{\langle \, \msf{y} \mid (\oper{D}^{-1})^\dag \oper{K}(\msf{x}) \mid \,\, }{\fluctw_0}{ \rangle \, \,  \langle \, }{ \fluctw_0}
				\contraction{\inner{ \msf{y} \mid (\oper{D}^{-1})^\dag \,	 \oper{K}(\msf{x}) \mid \fluctw_0} \langle \, \fluctw_0 \mid  \oper{K}(\msf{y}) \mid \,\,}{\fluctw_0 }{ \, \langle \, \,   \rangle \,  }{ \fluctw_0}
				\inner{ \msf{y} \mid (\oper{D}^{-1})^\dag \,	 \oper{K}(\msf{x}) \mid \fluctw_0} 
				\inner{ \fluctw_0 \mid \oper{K}^\dag(\msf{y}) \mid \fluctw_0}
				\bralong{\fluctw_0}
				\notag \\
		&			\quad \quad
			=	2	\bray \,	(\oper{D}^{-1})^\dag \, \oper{K}(\msf{x}) \,
					\overline{ \ketlong{\fluctw_0}  \bralong{\fluctw_0} }
					 \, \oper{K}^\dag(\msf{y}) \, 
					 \overline{ \ketlong{\fluctw_0}  \bralong{\fluctw_0} }
					\notag \\
		&			\quad \quad
			=	2	\bray 	\, (\oper{D}^{-1})^\dag\,  \oper{K}(\msf{x})  \, \oper{W}_0  \, \oper{K}^\dag(\msf{y}) \, \oper{W}_0 ,
\end{align}
where the connecting overlines indicate the correlations of the terms that contribute to the final result. [This notation is also commonly used in quantum field theory \citep{Peskin:2018}.] It is to be noted that this result can also be obtained by inserting the completeness relation $\widehat{1} = \int \mathrm{d}^3 \msf{x}' \ketlong{\msf{x}'} \bralong{\msf{x}'}$ and then using \Eq{eq:closure_quasi_quartic} explicitly. In the third line, we used the symmetry property given in  \Eq{eq:basic_symmetry}.  Substituting this result into \Eq{eq:closure_first} gives
\begin{equation}
	\overline{ \ketlong{ f_{\rm nl} [\widetilde{\phi} ,\fluctw_0 ] }  \bralong{\fluctw_0} }
		=		-\ui \int  \mathrm{d}^3 \msf{x} \,	\mathrm{d}^3 \msf{y} \,
							\ketx \bray
							(\oper{D}^{-1})^\dag\,
							\oper{K}(\msf{x}) \,
							\oper{W}_0 \,
							\oper{K}^\dag (\msf{y}) \,
							\oper{W}_0	.		
	\label{eq:closure_res_I}	
\end{equation}
In a similar manner, substituting \Eq{eq:closure_phi} into the third term in \Eq{eq:closure_main_aux} leads to
\begin{multline}
	 \overline{    \ketlong{ f_{\rm nl}  [\fluctw_0 ,\fluctw_0 ] } 	 \bralong{\widetilde{\phi}} }
			=	\int \mathrm{d}^3 \msf{x} \, \ketx \overline{ 
					\inner{ \fluctw_0 \mid \oper{K}(\msf{x}) \mid \fluctw_0}
					\bralong{\widetilde{\phi}} }
			=	- \frac{ \ui }{2}	 \int 	\mathrm{d}^3 \msf{x} \, \mathrm{d}^3 \msf{y} \, 	
							\ketx \bray  (\oper{D}^{-1})^\dag  \\
							\times
							\overline{	\inner{ \fluctw_0 \mid \oper{K}(\msf{x}) \mid \fluctw_0}
							\big[  \inner{ \fluctw_0 \mid \oper{K}^\dag (\msf{y}) \mid \fluctw_0} 
					 					-	\overline{ \inner{ \fluctw_0 \mid \oper{K}^\dag(\msf{y}) \mid
					 									 \fluctw_0} } \, \big] \, }.
	\label{eq:closure_second}				
\end{multline}
As before, when using the quasinormal approximation, we obtain
\begin{align}
	& \overline{ \inner{   \fluctw_0 \mid \oper{K}(\msf{x}) \mid \fluctw_0 } 
			\, \big[ \, \inner{ \fluctw_0 \mid \oper{K}^\dag (\msf{y}) \mid \fluctw_0} 
			-	\overline{ \inner{ \fluctw_0 \mid \oper{K}^\dag(\msf{y}) \mid \fluctw_0} } \, \big] \, \,  }
				\notag \\
		&	\quad \quad \quad
			=	\contraction{\langle \,\,}{\fluctw_0}{\mid \oper{K}(\msf{x}) \mid \fluctw_0 \rangle \langle \, \,\, }{ \fluctw_0}
				\contraction[2ex]{\langle \, \fluctw_0 \mid \oper{K}(\msf{x}) \mid \, }{\fluctw_0 }{ \, \rangle \,\,\,\, \langle \, \fluctw_0 \mid \oper{K}(\msf{y}) \mid }{ \fluctw_0}
				\inner{ \fluctw_0 \mid \oper{K}(\msf{x}) \mid \fluctw_0}
				\inner{ \fluctw_0 \mid \oper{K}^\dag (\msf{y}) \mid \fluctw_0} 
		+
				\contraction{\langle \, \fluctw_0 \mid \oper{K}(\msf{x}) \mid }{\fluctw_0 \, \, }{ \rangle \langle \, \,}{ \fluctw_0}
				\contraction[2ex]{\langle \, \,}{\fluctw_0 }{ \mid \oper{K}(\msf{x}) \mid \fluctw_0 \, \rangle   \,\,\,\, \langle \, \fluctw_0 \mid \oper{K}(\msf{y}) \mid }{ \fluctw_0}
				\inner{ \fluctw_0 \mid \oper{K}(\msf{x}) \mid \fluctw_0}
				\inner{ \fluctw_0 \mid \oper{K}^\dag (\msf{y}) \mid \fluctw_0} 
				\notag \\
		&	\quad \quad \quad
			= 2 \contraction{\langle \, \fluctw_0 \mid \oper{K}(\msf{x}) \mid }{\fluctw_0 \, \, }{ \rangle \langle \, \,}{ \fluctw_0}
				\contraction[2ex]{\langle \, \,}{\fluctw_0 }{ \mid \oper{K}(\msf{x}) \mid \fluctw_0 \, \rangle   \,\,\,\, \langle \, \fluctw_0 \mid \oper{K}(\msf{y}) \mid }{ \fluctw_0}
				\inner{ \fluctw_0 \mid \oper{K}(\msf{x}) \mid \fluctw_0}
				\inner{ \fluctw_0 \mid \oper{K}^\dag (\msf{y}) \mid \fluctw_0}
				\notag \\
		&	\quad \quad \quad
			=	2	\, \mathrm{Tr} [ \, \oper{K}(\msf{x})  \, \overline{ \ketlong{\fluctw_0} \bralong{\fluctw_0} } \, \oper{K}^\dag(\msf{y})  \,
											\overline{ \ketlong{\fluctw_0}  \bralong{\fluctw_0} }	\,]
				\notag \\
		&	\quad \quad \quad
			= 	2 \,	\mathrm{Tr} [ \, \oper{K}(\msf{x}) \, \oper{W}_0 \, \oper{K}^\dag(\msf{y}) \, \oper{W}_0 \, ] ,
	\label{eq:closure_calc_aux}	
\end{align}
where we introduced the trace operation $\smash{ \mathrm{Tr} \, \oper{A} \,  = \int \mathrm{d}^3 \msf{x} \inner{ \msf{x} \mid \oper{A} \mid \msf{x}} }$. By using the identity operator $\smash{ \widehat{1} = \int \mathrm{d}^3 \msf{x}' \ketlong{\msf{x}'} \bralong{\msf{x}'} }$, we are able to write $\inner{ w_0 \mid \oper{A} \mid w_0 } = \mathrm{Tr}( \, \oper{A} \ketlong{w_0} \bralong{w_0} \, )$. Also, note that the trace $\mathrm{Tr} [ \, \oper{K}(\msf{x}) \, \oper{W}_0 \, \oper{K}^\dag(\msf{y}) \, \oper{W}_0 \, ]$ is not an operator but rather a function of the spacetime coordinates $\msf{x}$ and $\msf{y}$. Upon inserting \Eq{eq:closure_calc_aux} into \Eq{eq:closure_second}, we obtain

\begin{equation}
	\overline{    \ketlong{ f_{\rm nl} [\fluctw_0 ,\fluctw_0 ] }  \bralong{\widetilde{\phi}} }
		=  - \ui \int 	\mathrm{d}^3 \msf{x} \, \mathrm{d}^3 \msf{y} \,
				 \ketx \bray (\oper{D}^{-1})^\dag  \,
				 \mathrm{Tr} [ \, \oper{K}(\msf{x}) \, \oper{W}_0 \, \oper{K}^\dag(\msf{y}) \, \oper{W}_0 \, ] .
		\label{eq:closure_res_II}		
\end{equation}
We then substitute \Eqs{eq:closure_res_I} and \eq{eq:closure_res_II} into \Eq{eq:closure_main_aux}. Approximating $\oper{W}_0 \simeq \oper{W}$, which is valid to the leading order of accuracy of the theory, gives a closed equation for the correlation operator:
\begin{align}
		[\oper{D}_{\rm H}, \oper{W} ]_- 	+ \ui  [\oper{D}_{\rm A}, \oper{W} ]_+ 
			=	&	\, 2 \ui \epnl^2	\big[ \oper{F} \,  (\oper{D}^{-1})^\dag \big]_{\rm A}  
					- 2 \ui \epnl^2 	\big[ \widehat{\eta} \, \oper{W} \big]_{\rm A} 
					+ 2 \ui \ep^2 	\big[  \oper{S} (\oper{D}^{-1})^\dag \big]_{\rm A} ,
		\label{eq:closure_main}
\end{align}
where
\begin{subequations}	\label{eq:closure_scattering}
\begin{gather}
	\widehat{\eta}	\doteq	-
									\int 	\mathrm{d}^3 \msf{x} \, 	\mathrm{d}^3 \msf{y} \, 	
									\ketx	\bray
									(\oper{D}^{-1})^\dag
									\oper{K}(\msf{x}) \, \oper{W} \, \oper{K}^\dag(\msf{y})   	, 
				\label{eq:closure_eta}			\\
	 \oper{F}		\doteq
	 								\frac{1}{2}
	 								\int 	\mathrm{d}^3 \msf{x} \, 	\mathrm{d}^3 \msf{y} \, 	
									\ketx\bray 
									\mathrm{Tr} [ \, \oper{K}(\msf{x}) \, \oper{W} \, \oper{K}^\dag (\msf{y}) \, \oper{W} \, ]  		.
				\label{eq:closure_F}		 
\end{gather}
\end{subequations}

In summary, \Eqs{eq:stats_mean}, \eq{eq:closure_main}, and \eq{eq:closure_scattering} form a complete set of equations for the zonal flow $U(t,y)$ and for the correlation operator $\smash{\oper{W}}$. Multiplying \Eq{eq:closure_main} by $\bralong{t,\vec{x}}$ and $\ketlong{t',\vec{x}'}$ leads to the ``quasinormal" equation for the two-point correlation function written in the double-physical coordinate space $(\msf{x}, \msf{x'})$. However, we shall instead project these equations into the phase space by using the Weyl transform. Then, by adopting approximations that are consistent with the GO description of eikonal fields, we shall obtain a WKE model describing the vorticity fluctuations. (Readers who are not familiar with the Weyl calculus are encouraged to read \App{app:weyl} before continuing further.)

\section{Dynamics in the ray phase space}
\label{sec:WKE}

\subsection{Wigner--Moyal equation}

The Weyl transform is a mapping from operators in a Hilbert space into functions of phase space \citep{Tracy:2014to}. In this work, the Weyl transform is defined as
\begin{equation}
	A(\msf{x} , \msf{k} ) 	\doteq	\msf{W}[ \, \oper{A} \, ] = \int 	\mathrm{d}^3 \msf{s} \,
									 	\ue^{\ui \msf{k} \cdot \msf{s} / \ep} 
									 	\inner{ \msf{x} + \tfrac{1}{2} \msf{s}  \mid \oper{A} \mid \msf{x} - \tfrac{1}{2} \msf{s} },
	\label{eq:wke_weyl}
\end{equation}
where $\msf{k} \doteq (\omega, \vec{k})$, $ \msf{s} \doteq(\tau,\vec{s}) $, $ \msf{k} \cdot \msf{s} = \omega \tau -  \vec{k} \cdot \vec{s}$, and $\mathrm{d}^3 \msf{s} \doteq \mathrm{d} \tau \, \mathrm{d}^2 \vec{s}$. Here the integrals span over $\mathbb{R}^3$. The \emph{Weyl symbol} $A(\msf{x} , \msf{k} ) $ of an operator $\smash{\oper{A}}$ is a function on the extended six-dimensional phase space $(\msf{x} , \msf{k} ) = (t,\vec{x},\omega, \vec{k})$; i.e., $A=A(t,\vec{x},\omega, \vec{k})$. Physically, $A(\msf{x} , \msf{k} )$ can be interpreted as a local Fourier transform of the spacetime representation of the operator $\smash{\oper{A}}$ [see, for instance, \Eq{eq:weyl_weyl_x_rep}]. 

The Weyl symbol $W(t,y,\omega,\vec{k})$ corresponding to $\smash{\oper{W}}$ is referred to as the \emph{Wigner function} of the vorticity fluctuations \citep{Wigner:1932cz}. Since $W(t,y,\omega,\vec{k})$ is a Weyl symbol of the zonal-averaged operator \eq{eq:stats_Wigner}, it does not depend on the $x$ coordinate. Upon following \App{app:zonal_average}, one can write $W(t,y,\omega,\vec{k})$ explicitly as
\begin{align}
	W(t,y,\omega,\vec{k}) 
		&	=	\int 	\mathrm{d}^3 \msf{s} \,
									\ue^{\ui \msf{k} \cdot \msf{s} / \ep } 
									\inner{ \msf{x} + \tfrac{1}{2} \msf{s} \mid \oper{W} \mid \msf{x} - \tfrac{1}{2} \msf{s} } \notag \\
		&	=	\int 	\mathrm{d}^3 \msf{s} \, \mathrm{d}x \, 
									\ue^{\ui \msf{k} \cdot \msf{s} / \ep} 
									\avg{	 	\fluctw( \msf{x} + \tfrac{1}{2} \msf{s} ) \,
												\fluctw( \msf{x} - \tfrac{1}{2} \msf{s}  ) } /L_x ,
\end{align}
Since $\fluctw$ is real, then $W(t,y,\omega,\vec{k})=W(t,y,-\omega,-\vec{k})$. Also, $W(t,y,\omega,\vec{k})$ is a real function because $\smash{\oper{W}}$ is Hermitian. Similar arguments apply to the Weyl symbol $S(t,y,\omega,\vec{k})$ corresponding to the operator $\smash{\oper{S} }$.

Applying the Weyl transform to \Eq{eq:closure_main} leads to the Wigner--Moyal formulation of DW--ZF dynamics:
\begin{align}
		\moysin{ D_{\rm H}, W }	+	&    \ui \, \moycos{D_{\rm A}, W} \notag \\
			=	& \, 2  \ui \epnl^2 \, \mathrm{Im} \left\{ F \star  [D^{-1}]^*  \right\}	 
				- 2 \ui \epnl^2 \, \mathrm{Im} \left( \eta \star  W \right)		 	
			    + 2 \ui  \ep^2 \, \mathrm{Im}  \left\{ S \star  [D^{-1}]^* \right\} .
		\label{eq:wke_wigner_moyal}
\end{align}
Here $D(t,y,\omega,\vec{k})$ is the Weyl symbol corresponding to $\smash{\oper{D}}$. From the properties of the Weyl transform, we observe that $D_{\rm H} = \mathrm{Re} \, D$ and $D_{\rm A}= \mathrm{Im} \, D$ are the (real) Weyl symbols corresponding to the operators $\smash{\oper{D}_{\rm H}}$ and $\smash{\oper{D}_{\rm A}}$, respectively. (``$\mathrm{Re}$" and ``$\mathrm{Im}$" denote the real and imaginary parts, respectively.) These are given by
\begin{subequations}	\label{eq:wke_symbol_D}
	\begin{align}
		D_{\rm H}(t,y,\omega,\vec{k}) 	
						&	= 	\omega - k_x U 
								+ \tfrac{1}{2} \moycos{  ( \beta - U'') ,  k_x /  k_{\rm D}^2 }, \\
		D_{\rm A}(t,y,\vec{k}) 				
						&	=	\ep \mu_{\rm dw}	
								- \tfrac{1}{2} \moysin{  U'' ,  k_x /  k_{\rm D}^2 } .
	\end{align}
\end{subequations}
In addition, $F(t,y,\omega,\vec{k})$, $\eta(t,y,\omega,\vec{k})$, and $[D^{-1}]^*(t,y,\omega,\vec{k})$ are the Weyl symbols corresponding to $\smash{\oper{F}}$, $\smash{\oper{\eta}}$, and $\smash{(\oper{D}^{-1})^\dag}$, respectively. (These will be calculated explicitly later.) The Moyal product \eq{eq:weyl_Moyal} is denoted by ``$\star$'', and the brackets $\moysin{ \cdot , \cdot }$ and $\moycos{\cdot , \cdot }$ are the Moyal brackets [\Eqs{eq:weyl_sine_bracket} and \eq{eq:weyl_cosine_bracket}]. Basic properties of the Moyal product and the Moyal brackets are given in \App{app:weyl}.

Modulo the statistical closure introduced in \Sec{sec:stats}, \Eq{eq:wke_wigner_moyal} is an exact equation for the dynamics of the Wigner function. However, \Eq{eq:wke_wigner_moyal} is difficult to solve as is, both analytically and numerically. Thus, we shall reduce \Eq{eq:wke_wigner_moyal} to a first-order partial-differential equation (PDE) in phase space by using the GO approximation.

\subsection{Collisional wave kinetic equation}
\label{sec:wke_derivation}

In order to simplify \Eq{eq:wke_wigner_moyal}, we shall expand the Moyal products and brackets in terms of the ordering parameter $\ep$. As a reminder, the parameter $\ep$ was introduced in \Eq{eq:GO_condition} in order to denote the spatio-temporal scale separation between the DW and ZF dynamics. From the definition of the Moyal product [\Eq{eq:weyl_Moyal2}], one has
\begin{equation}	
	A \star B 		=	A B 	+ \frac{\ui \ep}{2} ( A \overleftrightarrow{\mc{L}} B )  + O(\ep^2)	, 
		\label{eq:wke_product_approx}
\end{equation}
where $\overleftrightarrow{\mc{L}}$ is the Janus operator [\Eq{eq:weyl_poisson_braket}], which basically serves as the canonical Poisson bracket in the extended six-dimensional phase space $(t,\vec{x},\omega, \vec{k})$. Likewise, the Moyal brackets in \Eqs{eq:weyl_sine_bracket} and \eq{eq:weyl_cosine_bracket} are approximated by
\begin{equation}	\label{eq:wke_bracket_approx}
		\moysin{A , B} 	
							=	\ui \ep ( A \overleftrightarrow{\mc{L}} B )
								+ O(\ep^3)	, \quad \quad 		
		\moycos{A , B} 	
							=	2 A B + O(\ep^2) .
\end{equation}
For the purposes of this paper, higher-order corrections to \Eqs{eq:wke_product_approx} and \eq{eq:wke_bracket_approx} will not be needed. It is to be noted that the asymptotic expansions above are valid as long as the functions involved are smooth.

In addition to asymptotically expanding the Moyal products and brackets, let us find a proper ansatz for the Wigner function. Note that \Eq{eq:stats_eq_original} can be written as $\smash{\oper{D}_{\rm H} \oper{W} = O(\ep, \epnl)}$, where we assumed small dissipation $[D_{\rm A} \sim O(\epsilon)]$. In the Weyl representation, this equation becomes $D_{\rm H} \star W = O(\ep, \epnl)$. Using \Eq{eq:wke_product_approx} gives
\begin{equation}
	D_{\rm H} (t,y, \omega, \vec{k})  \, W(t,y,\omega, \vec{k}) 
			\simeq O(\ep, \epnl  ) .
	\label{eq:wke_dispersion}
\end{equation}
To satisfy this equation, we adopt the GO ansatz \citep{McDonald:1985ib,McDonald:1991kk}
\begin{equation}
	W(t,y,\omega,\vec{k}) 
		=  	2 \upi \ep 	\, 
			\delta \boldsymbol{(} D_{\rm H} (t,y, \omega, \vec{k}) \boldsymbol{)} \,
			J (t,y,\vec{k})	 .
	\label{eq:wke_ansatz}
\end{equation}
Here $J(t,y,\vec{k})$ is interpreted as the wave-action density for the vorticity fluctuations, and $2\upi \epsilon$ is added to ensure the proper normalization.

To remain consistent with the asymptotic expansion in \Eqs{eq:wke_product_approx} and \eq{eq:wke_bracket_approx}, the symbols $D_{\rm H}$ and $D_{\rm A}$ in \Eqs{eq:wke_symbol_D} are approximated as well to lowest order in $\ep$:
\begin{subequations}	\label{eq:wke_D_approx}
	\begin{align}
		D_{\rm H}(t,y,\omega,\vec{k}) 	
				&	\simeq	 	\omega - k_x U 
									+ ( \beta - U'') k_x /  k_{\rm D}^2 , 
			\label{eq:wke_DH}\\
		D_{\rm A}(t,y,\vec{k}) 
				&	\simeq		\ep \mu_{\rm dw}	+ \ep  k_x k_y U''' /  k_{\rm D}^4  ,
	\end{align}
\end{subequations}
where we substituted \Eqs{eq:wke_bracket_approx}.

Let us now obtain an approximate expression for $[D^{-1}]^*$ in \Eq{eq:wke_wigner_moyal}. As a reminder, $[D^{-1}]$ and $D^{-1}$ are not the equivalent; $[D^{-1}]$ is the Weyl symbol of $\smash{\oper{D}^{-1}}$ while $D^{-1}$ is simply the inverse of the Weyl symbol $D$ corresponding to $\oper{D}$. However, note that the Weyl representation of $\smash{\oper{D} \oper{D}^{-1} = \widehat{1}}$ is $D\star [ D^{-1} ]= 1$. By replacing the Moyal product with an ordinary product, we obtain $[D^{-1}] \simeq D^{-1}$. Then, by using the Sokhotski--Plemelj theorem, we replace $[D^{-1}]^* $ with its limiting form as $\ep$ tends to zero:
\begin{equation}
	[D^{-1}]^* 	\simeq 	\lim_{\ep\to 0} \, \frac{1}{D_{\rm H} - \ui D_{\rm A} } 
							= 	\ui \upi \delta(D_{\rm H}) 	+ \mc{P} \, \frac{1}{D_{\rm H}}  ,
	\label{eq:wke_D}
\end{equation}
where ``$\mc{P}$" denotes the Cauchy principal value.

We then insert \Eqs{eq:wke_bracket_approx}--\eq{eq:wke_D} into \Eq{eq:wke_wigner_moyal} and integrate over the frequency variable $\omega$. Afterwards, we expand the Moyal products and brackets using \Eqs{eq:wke_product_approx} and \eq{eq:wke_bracket_approx}. Although the GO ansatz \eq{eq:wke_ansatz} and the expression for $[D^{-1}]$ in \Eq{eq:wke_D} involve Dirac delta functions whose derivatives in phase space are not smooth, in \App{app:simplification} we show that these singularities can be removed via integration by parts. After some calculations, we obtain the following to lowest order in $\ep$:
\begin{multline}
	\ep  \int \mathrm{d} \omega \, 
			\left[  \ep ( D_{\rm H}  \overleftrightarrow{\mc{L}} J )  \, \delta(D_{\rm H} )  
					+ 2 D_{\rm A} \delta(D_{\rm H} ) J \right]   
			[1+ O(\ep)]  \\
		=	
			\epnl^2 \int \mathrm{d} \omega \, \left[ 
		 	 \,   F   \delta(D_{\rm H})	 -  2 \mathrm{Im} \left( \eta     \right)	J  \delta(D_{\rm H}) \right] [1+O(\ep)]  	
			 + \ep^2 \int \mathrm{d} \omega \, S \delta(D_{\rm H})
				 	[1+O(\ep)] .\\
	\label{eq:wke_aux}
\end{multline}
Note that in \Eq{eq:wke_aux} the Weyl symbols $W$, $F$, and $S$  are real because their associated operators are Hermitian.

For simplicity, we consider that the nonlinear scaling parameter $\epnl$ and the GO parameter $\ep$ scale as $\ep \sim \epnl$. Upon integrating \Eq{eq:wke_aux} over the frequency and neglecting higher-order terms in $\ep$, we obtain the \emph{collisional wave kinetic equation} (cWKE)
\begin{equation}
	\pd_t J + \{ J , \Omega \}	 
		=	- 2 \mu_{\rm dw} J 
			+ 2 \Gamma J 
			+ 	S_{\rm ext} 
			+	\ep \, C[J,J],
	\label{eq:wke}
\end{equation}
where $\{ \cdot , \cdot \}= \overleftarrow{\pd_\vec{x}} \cdot \overrightarrow{\pd_\vec{k}} - \overleftarrow{\pd_\vec{k}} \cdot \overrightarrow{\pd_\vec{x}}$. The wave frequency $\Omega$ and dissipation term $\Gamma$ are respectively obtained from the lowest-order expansion in $\ep$ of $D_{\rm H}$ and $D_{\rm A}$ in \Eqs{eq:wke_D_approx}:
\begin{subequations}	\label{eq:wke_omega_gamma}
	\begin{align}
		\Omega(t,y,\vec{k}) 	& \doteq 	k_x U - (\beta - U'') k_x /k_{\rm D}^2 ,  	
				\label{eq:wke_omega} \\
		\Gamma(t,y,\vec{k}) 	& \doteq 	- U'''  k_x k_y /  k_{\rm D}^4 .										
				\label{eq:wke_gamma}
	\end{align}
\end{subequations}

The external source term $S_{\rm ext}(t, y,\vec{k})$ appearing in \Eq{eq:wke} is given by
\begin{align}
	S_{\rm ext}(t, y,\vec{k})
		=	\int 	\mathrm{d}^3 \msf{s} \,  \mathrm{d} x   \, 
									\ue^{\ui \msf{k} \cdot \msf{s} / \ep } 
									\avg{	 	\fluctxi( \msf{x} + \tfrac{1}{2} \msf{s} ) \,
												\fluctxi( \msf{x} - \tfrac{1}{2} \msf{s}  ) } /L_x .
\end{align}
Assuming that $\widetilde{\xi}(t,\vec{x})$ is white noise leads to
\begin{gather}
	\int \mathrm{d} x \,  \avg{ \widetilde{\xi}(t,\vec{x}) \widetilde{\xi}(t',\vec{x}')} / L_x
			= \delta(t-t')\,\,\Xi \boldsymbol{(} \tfrac{1}{2}(y + y') , \vec{x}-\vec{x}' \boldsymbol{)}.
\end{gather}
Thus, the Weyl symbol of the zonal-averaged operator for the stochastic forcing is
\begin{align}
	S_{\rm ext}(t,y,\vec{k})	
		= 2 \int \mathrm{d}^2 \vec{s} \, \,
		      \Xi(y, \vec{s}) \cos(\vec{p} \cdot \vec{s} / \ep).
	\label{eq:wke_source}
\end{align}
Here we used $ \Xi(y, \vec{s}) =  \Xi(y, -\vec{s})$, which is due to the fact that $\widetilde{\xi}$ is real by definition.

The term $C[J,J](t,y,\vec{k})$ in \Eq{eq:wke} represents wave--wave collisions and is given by
\begin{equation}
	C[J,J](t, y,\vec{k})	\doteq 	S_{\rm nl}[J,J] - 2\gamma_{\rm nl}[J]  J .
	\label{eq:wke_scattering}
\end{equation}
As shown in \App{app:nonlinear}, to the leading order in $\ep$, $\gamma_{\rm nl}[J]  $ and $S_{\rm nl}[J,J]$ are given by
\begin{subequations}	\label{eq:wke_coeff_nl}
		\begin{align}
		\gamma_{\rm nl}[J] (t,y,\vec{k}) 
			&  \doteq	\int \frac{\mathrm{d}^2 \vec{p} \, \mathrm{d}^2 \vec{q} }{(2 \upi \ep )^2} \, 
						\delta^2(\vec{k} -\vec{p} -\vec{q}) \,
						\Theta(t,y,\vec{k},\vec{p},\vec{q}) \,
						M(\vec{p}, \vec{q}) M(\vec{p}, \vec{k}) \,
						J(t,y,\vec{p}) ,
						 \label{eq:wke_gamma_nl} \\
		S_{\rm nl}[J,J](t, y,\vec{k}) 
			&	\doteq	\int \frac{\mathrm{d}^2 \vec{p} \, \mathrm{d}^2 \vec{q} }{(2 \upi \ep )^2} \, 
						\delta^2(\vec{k} -\vec{p} -\vec{q}) \,
						\Theta(t,y,\vec{k},\vec{p},\vec{q})	 \, 
						|M(\vec{p}, \vec{q})|^2 \,
						J(t,y,\vec{p}) J(t,y,\vec{q}).
						 \label{eq:wke_source_nl} 	
		\end{align}
\end{subequations}
Here $\Theta(t,y,\vec{k},\vec{p},\vec{q}) \doteq \upi \delta (\Delta \Omega)$, and
\begin{equation}
	\Delta\Omega(t,y,\vec{k},\vec{p},\vec{q}) \doteq \Omega(t,y,\vec{k})  - \Omega(t,y,\vec{p}) - \Omega(t,y,\vec{q}).
	\label{eq:wke_resonance}
\end{equation}
One can identify $\Delta \Omega=0$ as the frequency-resonance condition. Finally, the kernel $M(\vec{p},\vec{q})$ in \Eqs{eq:wke_coeff_nl} is
\begin{equation}
	M(\vec{p},\vec{q})
		\doteq \vec{e}_z \cdot (\vec{p}\times \vec{q}) 
			\left( q_{\rm D}^{-2} -  p_{\rm D}^{-2} \right).
	\label{eq:wke_M}
\end{equation}

\subsection{Dynamics of the ZF velocity $U$}

Returning to \Eq{eq:Dirac_mean} for the zonal flow velocity, the term on the right-hand side can be rewritten in terms of the Wigner function $W$ by using \Eq{eq:weyl_trace}. One has
\begin{align}
	\pd_t U + \mu_{\rm zf} U
		& 	=		\epnl^2
					\frac{\pd}{\pd y} 
					\int \frac{\mathrm{d} \omega \, \mathrm{d}^2 \vec{k} }{(2 \upi \ep )^3} \,
					\frac{k_x}{ k_{\rm D}^2} \star 
					W(t,y,\omega,\vec{k})
					\star \frac{k_y}{k_{\rm D}^2} , \notag \\
		& 	=		\epnl^2
					\frac{\pd}{\pd y} 
					\int \frac{ \mathrm{d}^2 \vec{k} }{(2 \upi \ep )^2} \,
					\frac{k_x}{ k_{\rm D}^2} \star 
					J(t,y,\vec{k})
					\star \frac{k_y}{k_{\rm D}^2} , 
\end{align}
where we used the Moyal product and substituted \Eq{eq:wke_ansatz}. Upon following \citet{Ruiz:2016gv}, we substitute $\ep\sim\epnl$ and obtain to lowest order in $\ep$:
\begin{equation}
	\pd_t U + \mu_{\rm zf} U = \ep^2
	\frac{\pd}{\pd y} 
	\int \frac{\mathrm{d}^2 \vec{k}}{(2 \upi \ep )^2} \,
		\frac{k_x k_y}{ k_{\rm D}^4} 
		J(t,y,\vec{k}) .
	\label{eq:wke_zonal}
\end{equation}
%

\section{Discussion}
\label{sec:discussion}

\subsection{Main equations}

Equations \eq{eq:wke} and \eq{eq:wke_zonal}, as well as \Eqs{eq:wke_omega_gamma}, \eq{eq:wke_scattering}, and \eq{eq:wke_coeff_nl}, are the main result of our work. These equations describe the coupled interaction between an \emph{incoherent} wave bath of DWs and a \emph{coherent} ZF velocity field. The cWKE \eq{eq:wke} governs the dynamics of the wave-action density $J$ for DWs. The left-hand side of \Eq{eq:wke} describes the wave refraction governed by the wave frequency $\Omega$ [\Eq{eq:wke_omega}], which serves as a Hamiltonian for the system. On the right-hand side, $\mu_{\rm dw}$ represents weak dissipation due to the external environment, and $\Gamma$ denotes linear dissipation caused by the ZFs \citep{Parker:2018eda,Ruiz:2016gv}. The term $S_{\rm ext}$ represents an external source term for the DW fluctuations.

The nonlinear term $C[J,J]$ in \Eq{eq:wke} plays the role of a \emph{wave scattering operator}. It is composed of two terms, $\gamma_{\rm nl}$ and $S_{\rm nl}$, which arise from nonlinear wave--wave interactions. The nonlinear source term $S_{\rm nl}$ in \eq{eq:wke_source_nl} is a bilinear functional on the action density $J$. It is always positive and represents contributions to $J$ coming from waves with wavevectors $\vec{p}$ and $\vec{q}$ different from $\vec{k}$. This term is also known as (the variance of) \emph{incoherent noise} \citep{Krommes:2002hva}. The nonlinear damping-rate term $\gamma_{\rm nl}$ in \Eq{eq:wke_gamma_nl} linearly depends on $J$ and represents a sink term where the wave action in the $\vec{k}$ wavevector is transferred to other modes with different wavevectors. The effects described by $\gamma_{\rm nl}$ are called the \emph{coherent response} \citep{Krommes:2002hva}. The other terms in \Eqs{eq:wke} and \eq{eq:wke_zonal} are the same as in the QL theory, and their physical meaning was already discussed elsewhere \citep{Parker:2016eu,Ruiz:2016gv,Zhu:2018fd,Parker:2018eda,Zhu:2018hk, Zhu:2018gk}.

\subsection{Conservation properties}
\label{sec:conservation}

Equations \eq{eq:wke} and \eq{eq:wke_zonal} inherit the same conservation laws of the original gHME \eq{eq:basic_hm}. In other words, for isolated systems ($S_{\rm ext}=0$ and $\mu_{\rm zf, dw}=0$), \Eqs{eq:wke} and \eq{eq:wke_zonal} conserve the total enstrophy and total energy
\begin{equation}
	\mc{Z}= \mc{Z}_{\rm zf} + \ep^2 \mc{Z}_{\rm dw} , 
		\quad \quad 
	\mc{E}= \mc{E}_{\rm zf} + \ep^2 \mc{E}_{\rm dw},
	\label{eq:conservation_main}
\end{equation}
where we used $\ep \sim \epnl$. Also, the expressions for the DW and ZF components of the enstrophy and energy are
\begin{subequations}	\label{eq:conservation_defs}
	\begin{align}
			\mc{Z}_{\rm zf} 	
					&	\doteq 	\frac{1}{2} \int \mathrm{d} y \, \overline{w}^2 
						=			\frac{1}{2} \int \mathrm{d} y \, (U')^2  ,  \\		
			\mc{Z}_{\rm dw} 	
					& \doteq 	\frac{1}{2} \int \mathrm{d}^2 \vec{x} \, \avg{ \fluctw^2 } 
						=	\frac{1}{2} \int \frac{\mathrm{d} y \, \mathrm{d}^2 \vec{k}}{(2 \upi \ep)^2} \,
								 J,   \\
			\mc{E}_{\rm zf} 	
					&	\doteq 	- \frac{1}{2} \int \mathrm{d} y \, \overline{w}\overline{\psi} 
						= 			\frac{1}{2} \int \mathrm{d} y \, U^2,  \\
			\mc{E}_{\rm dw} 	
					& \doteq 	- \frac{1}{2} \int \mathrm{d}^2 \vec{x} \, \avg{ \fluctw \fluctpsi } 
						=	\frac{1}{2} \int \frac{\mathrm{d} y \, \mathrm{d}^2 \vec{k}}{(2 \upi \ep)^2} \, 
								\frac{J}{k_{\rm D}^2} .
	\end{align}
\end{subequations}
These conservation laws are proven in \App{app:conservation}.

\subsection{Comparison with quasilinear models and weak turbulence theory}
\label{sec:comparison}

There exists a vast literature of QL WKE-based models of DW--ZF interactions. These models, which are called ``improved WKE'' (iWKE) by  \citet{Zhu:2018hk, Zhu:2018gk} and ``CE2-GO'' by \citet{Parker:2016eu,Parker:2018eda}, neglect wave--wave collisions and only differ from our derived model by letting $C[J,J]=0$. Earlier works also reported a simpler QL WKE \citep{Diamond:2005br,Smolyakov:1999jk,Smolyakov:2000be,Malkov:2001kp,Malkov:2001hv,Diamond:1994fd,Kim:2003jf,Kaw:2002ku,Trines:2005in,Singh:2014bh} with $\Omega = k_xU - \beta k_y / k_{\rm D}^2 $ and $\Gamma = 0$ . This simpler WKE-based model is referred as the ``traditional WKE'' (tWKE) in \citet{Ruiz:2016gv}, \citet{Zhu:2018fd,Zhu:2018hk, Zhu:2018gk}, and \citet{Parker:2018eda}. It is also referred as ``WKE" in \citet{Parker:2016eu}. It was shown in \citet{Parker:2016eu} that the tWKE does not conserve the total enstrophy (in contradiction with the underlying gHME model) and also leads to unphysical growth rates of the zonostrophic instability.  In contrast, the iWKE conserves total enstrophy and total energy and shows good qualitative agreement with direct numerical simulations of the QL gHME at least in certain regimes \citep{Parker:2016eu,Parker:2018eda}. Also, the iWKE was useful to obtain clarifying insights in the structure of the DW phase space and its relation to the Rayleigh--Kuo parameter $U''/\beta$ \citep{Zhu:2018gk}. For a more detailed discussion on the iWKE, see \citet{Parker:2018eda} and \citet{Zhu:2018fd}.

The WKE is based on the assumption of a scale separation between the DW and ZF dynamics. There also exist QL statistical models that do not assume scale separation. One notable example is the second-order quasilinear expansion, or CE2 \citep{Farrell:2003dm, Farrell:2007fq, Marston:2008gx, Srinivasan:2012im, AitChaalal:2016jx}, whose applications to DW-ZF physics were pursued in \citet{Farrell:2009ke}, \citet{Parker:2013hy,Parker:2014fc}, and \citet{Parker:2014tb}. Another QL theory is the Wigner--Moyal (WM) formulation \citep{Ruiz:2016gv,Zhu:2018fd} which describes the DW dynamics in ray phase space. Both the CE2 and WM theories are mathematically equivalent. However, the CE2 is based on the double-physical space representation (\Sec{sec:QL}) so it is not very intuitive. For the same reason, its robustness with respect to further approximations remains obscure. The WM formulation involves an infinite-order PDE in phase space, which arguably makes it more tractable.

It is worth mentioning that DW scattering has been widely studied in the context of homogeneous weak turbulence theory (WTT). [For recent reviews, see, \eg \citet{Krommes:2002hva}, \citet{Nazarenko:2011cr}, and \citet{Connaughton:2015kk}.] Previous works in WTT have derived wave--wave collision operators for DWs. However, in these models, both the DW and ZF components of the fields were considered as incoherent. In other words, the quasinormal approximation, or the related random phase approximation, where applied to both the mean and fluctuating quantities. However, direct statistical simulations based the QL approximation have shown that ZFs are not incoherent \citep{Farrell:2009ke,Srinivasan:2012im,Tobias:2013hk,Parker:2013hy,Parker:2014fc}. In light of this, \Eqs{eq:wke} and \eq{eq:wke_zonal} make a distinction between the statistics of the DWs and the ZF velocity field: the DW component is modeled as an incoherent wave ensemble (or wave bath) described by the cWKE, while ZFs are treated as coherent structures. Such partitioning of the statistics is expected to lead to different results.

In the absence of ZFs ($U = 0$), the collision operator $C[J,J]$ in \Eq{eq:wke_scattering} coincides with those derived previously within homogeneous WTT [see, \eg \citet{Connaughton:2015kk}]. Hence, the cWKE \eq{eq:wke} is expected to yield the same stationary, homogeneous, non-equilibrium solutions found in WTT. In other words, one would obtain the well-known Kolmogorov--Zakharov spectra for DW turbulence. In this work, we shall not further discuss this topic. The interested reader can refer to, \eg Sec. 3.3 of \citet{Connaughton:2015kk}. 

Upon using the wave--wave collision operator that we propose here, future work will be devoted to studying the possible modifications to the DW--ZF dynamics caused by $C[J,J]$ at nonzero $U$. In particular, the frequency-resonance condition in WTT depends on a simplified expression for the DW frequency: $\Omega(t,y,\vec{k}) \simeq -\beta k_x / k_{\rm D}^2$, which neglects the ZF velocity. Hence, one has
\begin{equation*}
	\Theta_{\rm WTT}(t,y,\vec{k},\vec{p},\vec{q})
		=	\frac{\upi}{\beta } \,
						\delta \left( 
							\frac{k_x}{k_{\rm D}^2}
								-\frac{p_x}{p_{\rm D}^2}
								-\frac{q_x}{q_{\rm D}^2} \right).
\end{equation*}
In contrast, in the cWKE,  $\Omega$ is given by \Eq{eq:wke_omega}, so
\begin{equation}
	\Theta(t,y,\vec{k},\vec{p},\vec{q})
		=	\frac{\upi}{|\beta - U''(t,y) |} \,
						\delta \left( 
							\frac{k_x}{k_{\rm D}^2}
								-\frac{p_x}{p_{\rm D}^2}
								-\frac{q_x}{q_{\rm D}^2} \right),
\end{equation}
where the contribution of the Doppler shifts canceled out due to the spatial-resonance condition $(\vec{k}=\vec{p}+\vec{q})$. The factor $|\beta - U''|$ appearing in the denominator is related to the Rayleigh--Kuo threshold. The relevance of this threshold to the GO dynamics of DWs was shown by \cite{Zhu:2018gk}, and in a broader context, it also marks the onset of the tertiary instability \citep{Zhu:2018fd,Kuo:1949cr,Numata:2007ek}. From this result, it seems that the present theory breaks down in regions where $\beta-U'' \simeq \ep$. In such regions, the DW wave frequency tends to zero, and the interaction time between the waves becomes long. In general, WTT is based on the assumption that the interaction time between the waves are small, so this violates the assumed orderings used to close the equations and leads to the breakdown of the model.

\subsection{Comparison with the generalized quasilinear approximation}

Let us also mention that the present work is somewhat similar in spirit to the \emph{generalized quasilinear (GQL) approximation} proposed by \citet{Marston:2016ff}, where the dynamical fields are separated into small and large zonal scales via a spectral filter that depends on a wavenumber cutoff. In order to incorporate the nonlinear energy transfer due to eddy--eddy interactions, \citet{Marston:2016ff} includes nonlinear large-scale ZF--ZF interactions but neglects small-scale DW--DW interactions. In simulations of zonal-jet formation \citep{Marston:2016ff} and rotating three-dimensional Couette flow \citep{Tobias:2016kp}, it was shown that the GQL gives an improvement in accuracy on calculations of the mean flows, spectra, and two-point correlation functions over models that are QL.

However, the GQL framework is not the same as the one presented here, namely, due to the following. First, in our work we decompose the fields into mean (i.e., independent of $x$) and fluctuating quantities. Within the gHME, nonlinear large-scale ZF--ZF interactions do not appear when using such decomposition. Second, we allow for small-scale DW--DW interactions. In consequence, the governing equation for the fluctuations remains nonlinear. Studying the differences between these models and developing a new framework that takes advantages of each model could be a topic of future research.

\subsection{Realizability of the cWKE model}
\label{sec:realizability}

Since it is known that the quasinormal approximation can lead to unrealizable statistics for the Euler equations \citep{Ogura:2006ck,Leslie:1973kv}, one may wonder if such issues regarding realizability could present themselves in the cWKE model proposed here. We anticipate that this is not the case for the following two reasons. First, the turbulence that we address is of different nature. Unlike the isotropic turbulence governed by the Euler equations, which is dominated by eddy--eddy interactions, the turbulence considered here is mainly determined by \emph{collective} wave--wave interactions. Second, as mentioned in \S\ref{sec:comparison}, the cWKE model is identical to homogeneous WTT theory \citep{Krommes:2002hva,Nazarenko:2011cr,Connaughton:2015kk} when no ZFs are present ($U = 0$). In homogeneous WTT, both the quasinormal approximation and the random-phase approximation lead to the same wave--wave scattering operators and show realizable statistics. Since the structures of the equations and of the collisional operator in our cWKE model are the same as those found in homogeneous WTT, we believe that the statistical cWKE model proposed here leads to meaningful statistics. However, this remains to be checked.

\section{Conclusions}
\label{sec:conclusions}

In this work, we present a nonlinear wave kinetic equation for studying DW--ZF interactions. In contrast with previous works that used the quasilinear approximation, here we perturbatively include nonlinear wave--wave collisions. Our derivation makes use of the Weyl calculus, in conjunction with the quasinormal statistical closure and the well-known geometrical-optics assumptions. The obtained model is similar to previous reported works \citep{Parker:2018eda,Ruiz:2016gv,Zhu:2018gk,Parker:2016eu,Zhu:2018hk} but also includes a nonlinear term describing wave--wave scattering. Unlike in WTT, the collision operator depends on the local ZF velocity and breaks down at the Rayleigh--Kuo threshold. Our model conserves both the total enstrophy and the total energy of the system.

The model presented here might allow us to investigate the effects of wave--wave scattering on the DW--ZF system. Several questions that could be addressed are the following. How will the spontaneous emergence of ZFs (also called the zonostrophic instability) be modified in the presence of wave--wave scattering? Regarding the saturated states, how will the partition of enstrophy and energy between DWs and ZFs change? Will the stationary Kolmogorov--Zakharov spectra for DWs be modified in the presence of ZFs? These and other questions will be subject to future investigations.

The authors thank H. Zhu, Y. Zhou, and J. B. Parker for helpful discussions. This work was supported by the U.S. DOE through Contract DE-AC02-09CH11466 and by Sandia National Laboratories. Sandia National Laboratories is a multimission laboratory managed and operated by National Technology and Engineering Solutions of Sandia, LLC., a wholly owned subsidiary of Honeywell International, Inc., for the U.S. DOE National Nuclear Security Administration under contract DE-NA-0003525. This paper describes objective technical results and analysis. Any subjective views or opinions that might be expressed in the paper do not necessarily represent the views of the U.S. DOE or the U.S. Government.

\appendix

\section{Conventions and definitions}
\label{app:conventions}


\subsection{Weyl calculus}
\label{app:weyl}

This appendix summarizes our conventions used for the Weyl calculus. For more information, see the reviews by \citet{Ruiz:2017wc}, \citet{Tracy:2014to}, \citet{Imre:1967fr}, \citet{BakerJr:1958bo}, and \citet{McDonald:1988dp}.

\par Let $\oper{A}$ be an operator defined on the Hilbert space $L^2(\mathbb{R}^3)$ with the inner product \eq{eq:abstract_inner}. The Weyl symbol $A(\msf{x} , \msf{k})$ is defined as the Weyl transform $\msf{W}[ \oper{A} ]$ of $\oper{A}$; namely,
\begin{equation}
	A(\msf{x} , \msf{k}) \doteq \msf{W} [ \, \oper{A}	\, ] =
				\int \mathrm{d}^3 \msf{s} \, \ue^{\ui \msf{k} \cdot \msf{s} / \ep }
				\inner{ \msf{x}+ \tfrac{1}{2}\msf{s} \mid \oper{A}	\mid \msf{x}	-	\tfrac{1}{2} \msf{s}	}  ,
	\label{eq:weyl_weyl_symbol}
\end{equation}
where $\ketx$ are the eigenstates of the position operator. Also, $\msf{k} \doteq (\omega, \vec{k})$, $ \msf{s} \doteq(\tau,\vec{s}) $, $ \msf{k} \cdot \msf{s} = \omega \tau -  \vec{k} \cdot \vec{s}$, and $\mathrm{d}^3 \msf{s} \doteq \mathrm{d} \tau \, \mathrm{d}^2 \vec{s}$. The integrals span $\mathbb{R}^3$. This description of the operators is known as the \emph{phase-space representation} since the Weyl symbols are functions of the six-dimensional phase space. Conversely, the inverse Weyl transform is given by
\begin{equation}
	\oper{A} = 
		\int	\frac{\mathrm{d}^3 \msf{x} \,  \mathrm{d}^3 \msf{k} \,  	\mathrm{d}^3 \msf{s} }{(2 \upi \ep)^3} \,
					\ue^{\ui \msf{k} \cdot \msf{s}/ \ep } 
					A( \msf{x} , \msf{k} ) \ketlong{ \msf{x} -\tfrac{1}{2} \msf{s} } \bralong{ \msf{x} + \tfrac{1}{2}\msf{s} },
	\label{eq:weyl_weyl_inverse}
\end{equation}
where $\mathrm{d}^3 \msf{k} \doteq \mathrm{d}\omega \, \mathrm{d}^2 \vec{k}$. The coordinate representation $\mcu{A}(\msf{x} , \msf{x}') = \inner{ \msf{x} \mid \oper{A}\mid \msf{x}' }$ of $\oper{A}$ is
\begin{equation}
	\mcu{A}(\msf{x} , \msf{x}') =  
			\int	\frac{\mathrm{d}^3 \msf{k} }{(2 \upi \ep )^3} \,
					\ue^{-\ui \msf{k} \cdot (\msf{x} - \msf{x}')/ \ep  }
					A		\left(	\frac{ \msf{x} + \msf{x}' }{2},	\msf{k}	\right)  .
	\label{eq:weyl_weyl_x_rep}
\end{equation}

In the following, we shall outline a number of useful properties of the Weyl transform.

\noindent (i) The trace $\mathrm{Tr} \, \oper{A}\, \doteq \int \mathrm{d}^3 \msf{x} \, \inner{ \msf{x} \mid  \oper{A} \mid  \msf{x} }$ of $\oper{A}$ is
\begin{equation}
	\mathrm{Tr} \, \oper{A} \, 
				=  \int \frac{\mathrm{d}^3\msf{x} \,  \mathrm{d}^3\msf{k} }{(2 \upi \ep)^3}\,  
					A( \msf{x} , \msf{k} ).
	\label{eq:weyl_trace}
\end{equation}

\noindent (ii) If $A( \msf{x} , \msf{k} )$ is the Weyl symbol of $\oper{A}$, then $A^*( \msf{x} , \msf{k} )$ is the Weyl symbol of $\oper{A}^\dag$. As a corollary, if the operator $\oper{A}$ is Hermitian $(\oper{A} = \oper{A}^\dag)$, then $A( \msf{x} , \msf{k} )$ is real.

\noindent (iii) For linear operators $\oper{A}$, $\oper{B}$ and $\oper{C}$ where $\oper{C} =\oper{A} \oper{B}$, the corresponding Weyl symbols satisfy
\begin{equation}
	C( \msf{x} , \msf{k} ) = A( \msf{x} , \msf{k} ) \star B( \msf{x} , \msf{k} ).
	\label{eq:weyl_Moyal}
\end{equation}
Here ``$\star$" refers to the \emph{Moyal product} \citep{Moyal:1949gj}, which is
\begin{equation}
		A( \msf{x} , \msf{k} ) \star B( \msf{x} , \msf{k} )
		\doteq 
		A( \msf{x} , \msf{k} )  \, \exp	\bigg(	\frac{\ui \ep}{2}	\overleftrightarrow{ \mc{L}}	\bigg) \, B( \msf{x} , \msf{k} ) .
	\label{eq:weyl_Moyal2}
\end{equation}
Also, $\overleftrightarrow{\mc{L}}$ is the \emph{Janus operator}
\begin{equation}
	\overleftrightarrow{\mc{L}} =
			\frac{\overleftarrow{\pd} }{\pd \vec{x}}  \cdot \frac{ \overrightarrow{\pd} }{\pd \vec{k} }
	-		\frac{\overleftarrow{\pd} }{\pd \vec{k}}  \cdot \frac{ \overrightarrow{\pd} }{\pd \vec{x} }	
	+		\frac{\overleftarrow{\pd} }{\pd \omega}  \frac{ \overrightarrow{\pd} }{\pd t}
	-		\frac{\overleftarrow{\pd} }{\pd t}  \frac{ \overrightarrow{\pd} }{\pd \omega}.
	\label{eq:weyl_poisson_braket}
\end{equation}
The arrows indicate the direction in which the derivatives act. Note that $A \overleftrightarrow{\mc{L}}B $ serves as the canonical Poisson bracket in the extended six-dimensional phase space $(t,\vec{x},\omega, \vec{k})$.

\noindent (iv) The Moyal product is associative; i.e., for arbitrary symbols $A$, $B$, and $C$, one has
\begin{equation}
	A \star B \star C 
			= (A \star B) \star C 
			= A \star (B \star C).
	\label{eq:weyl_associative}
\end{equation}

\noindent (v) The anti-symmetrized Moyal product defines the so-called \emph{Moyal bracket}, namely,
\begin{equation}
	\moysin{ A ,	B } 
	 		\doteq -i	\left(  A \star B - B \star A 	\right)
	 					= 	2 A \sin \bigg( \frac{ \ep }{2} \overleftrightarrow{ \mc{L}} \bigg) B .
	\label{eq:weyl_sine_bracket}
\end{equation}
Likewise, the symmetrized Moyal product is defined as
\begin{gather}
	 \moycos{ A , B} \doteq A \star B + B \star A 
	 						=	2 A \cos	\bigg(	\frac{\ep}{2}	\overleftrightarrow{ \mc{L}}	\bigg)  B.
 	\label{eq:weyl_cosine_bracket}
\end{gather}

\noindent (vi) For fields that vanish rapidly enough at infinity, when integrated over all phase space, the Moyal product of two symbols equals the regular product; i.e.,
\begin{gather}
 	\int \mathrm{d}^3 \msf{x} \, \mathrm{d}^3 \msf{k} \, 	A \star B 
   		= \int \mathrm{d}^3 \msf{x} \, \mathrm{d}^3 \msf{k} \, 		A B . 
 	\label{eq:weyl_int_everywhere}
\end{gather}

\noindent (vii) Now we tabulate some Weyl transforms of various operators. First, the Weyl transform of the identity is
\begin{equation}
	\msf{W} [ \, \widehat{1}	\, ] =	1.
\end{equation}
The Weyl transforms of the time and position operators are given by
\begin{equation}
	\msf{W} [ \, \widehat{t}\, ] =	t, 		 \quad	\quad	\msf{W} [ \, \widehat{\vec{x}}  ] = \vec{x},
\end{equation}
where $\widehat{t} = t $ and $\widehat{\vec{x}}= \vec{x}$ in the coordinate representation. Likewise,
\begin{equation}
	\msf{W} [ \, \widehat{\omega}  ] =	\omega, 		 \quad	\quad	\msf{W} [ \, \widehat{\vec{k}} ] = \vec{k},
\end{equation}
where $\widehat{\omega} = \ui \ep \pd_t $ and $\widehat{\vec{k}}= - \ui \ep \del$ in the coordinate representation. For any two operators $f(\widehat{t}, \widehat{\vec{x}})$ and $g(\widehat{\omega},\widehat{\vec{k}})$,
\begin{equation}
	\msf{W} [\, f(\widehat{t}, \widehat{\vec{x}}) 	\, ]	=   f(t,\vec{x}),  \quad	\quad
	\msf{W} [\, g(\widehat{\omega},\widehat{\vec{k}}) \,]=	 g(\omega, \vec{k}).
\end{equation}
Upon using the Moyal product \eq{eq:weyl_Moyal2}, one has
\begin{gather}
	\msf{W} [\,	\widehat{\vec{k}}	f(\widehat{\vec{x}}) 	\, ]	
								= \vec{k} f(\vec{x})	- \tfrac{\ui \ep}{2} \del 	f(\vec{x}), \\
	\msf{W} [\, 	f(\widehat{\vec{x}}) \widehat{\vec{k}} \, ]	
								= \vec{k} f(\vec{x})	+ \tfrac{ \ui \ep}{2} \del 	f(\vec{x}).
\end{gather}


\subsection{Zonal average of operators}
\label{app:zonal_average}

\begin{figure}
	\begin{center}
		\includegraphics[scale=1.1]{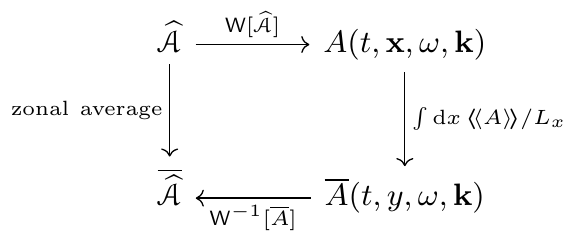}
	\end{center}
	\caption{Commutative diagram showing the definition of the zonal average of an operator $\oper{A}$.}
	\label{fig:diagram_zonal}
\end{figure}

The zonal average $\overline{\oper{A}}$ of any given operator $\oper{A}$ is defined through the Weyl calculus (\App{app:weyl}) and is schematically shown in \Fig{fig:diagram_zonal}. First, by using the Weyl transform \eq{eq:weyl_weyl_symbol}, one calculates the Weyl symbol $A(t,\vec{x},\omega,\vec{k})$ [\Eq{eq:weyl_weyl_symbol}] corresponding to the operator $\smash{\oper{A}}$. Then, one calculates the zonal average defined in \Sec{sec:mean_fluct} on the Weyl symbol $A(t,\vec{x},\omega,\vec{k})$. This leads to 
\begin{equation}
	\overline{A}(t,y,\omega, \vec{k}) \doteq \int \mathrm{d} x \, \avg{A(t,\vec{x},\omega,\vec{k})} / L_x.
\end{equation}
The zonal-averaged operator $\smash{\overline{\oper{A}}}$ is obtained by applying the inverse Weyl transform \eq{eq:weyl_weyl_inverse} on $\overline{A}(t,y,\omega, \vec{k})$.

From \Eq{eq:weyl_weyl_x_rep}, the coordinate representation of $\overline{\oper{A}}$ is
\begin{align}
	\overline{\mcu{A}} (t,\vec{x},t',\vec{x}') 
		&	=	\int	\frac{ \mathrm{d}\omega \, \mathrm{d}^2 \vec{k} \, }{(2 \upi \ep)^3} \,
					\ue^{-\ui \omega (t-t')/\ep + \ui \vec{k} \cdot (\vec{x} - \vec{x}' )/\ep } \, 
             \overline{A} \left(	\frac{t+t'}{2},\frac{y+y'}{2},\omega, \vec{k} \right) \notag \\
		&	\equiv	\overline{A}_{\rm F} \left(	\overline{t},\overline{y},\Delta t, \Delta \vec{x} \right),
\end{align}
where $\overline{A}_{\rm F}$ is the inverse Fourier transform on the frequency and wavevector variables of the Weyl symbol $\overline{A}(t,y,\omega, \vec{k})$. Also, $\overline{t}\doteq (t+t')/2$, $\overline{y}\doteq (y+y')/2$, $\Delta t \doteq t-t'$, and $\Delta \vec{x}\doteq \vec{x}-\vec{x}'$.

Another useful identity that we shall use is
\begin{equation}
	\overline{ \inner{ \msf{x} \mid \oper{A} \mid \msf{x} } } 
			=	\int \frac{\mathrm{d} x}{L_x} \avg{ \int \frac{\mathrm{d}^3 \msf{k}}{(2\upi \ep )^3} A(t,\vec{x},\omega , \vec{k} ) } 	
			= 	\int \frac{\mathrm{d}^3 \msf{k}}{(2\upi \ep)^3} \overline{A} (t, y , \omega , \vec{k} ) 
			=	\inner{ \msf{x} \mid \overline{\oper{A}} \mid \msf{x} },
	\label{eq:zonal_avg_identity}
\end{equation}
where we substituted \Eq{eq:weyl_weyl_x_rep} in the first and last lines.

\section{Auxiliary calculations}
\label{app:auxiliary}

\subsection{Simplifying the Moyal products}
\label{app:simplification}

To derive the WKE \eq{eq:wke}, we need to approximate the Moyal products appearing in \Eq{eq:wke_wigner_moyal}. The most difficult terms to approximate are those involving derivatives of Dirac delta functions. As an example, in this appendix we calculate the integral
\begin{equation}
	\mc{I} = \int \mathrm{d}\omega	\, A(z) \star W(z),
	\label{simp:eq:integral_orig}
\end{equation}
where $A(z)$ is an arbitrary function, $W(z)= 2 \upi \ep \delta \boldsymbol{(} D_{\rm H} (z) \boldsymbol{)} J(t,\vec{x},\vec{k})$ is the GO ansatz \eq{eq:wke_ansatz}, and $z\doteq(t,\vec{x},\omega, \vec{k})$. From \Eqs{eq:wke_DH} and \eq{eq:wke_omega}, the GO dispersion relation is $D_{\rm H}(z) \simeq \omega- \Omega(t,\vec{x},\vec{k})=0$. Substituting \Eq{eq:wke_ansatz} into \Eq{simp:eq:integral_orig} leads to
\begin{align}
	\mc{I} 
		&	= 2 \upi \ep \int \mathrm{d}\omega \,  A(z) \star [\delta(D_{\rm H}) J] \notag \\
		&	= 2 \upi \ep \int \mathrm{d}\omega \, A(z) \exp\left( \frac{ \ui \ep}{2} \overleftrightarrow{\mc{L}} \right)  [\delta(D_{\rm H}) J] 
		    = 2 \upi \ep \sum_{n=0}^\infty \frac{1}{n!} \left(\frac{ \ui \ep}{2}\right)^n \mc{I}_n, 
	\label{simp:eq:integral}
\end{align}
where $\overleftrightarrow{\mc{L}}$ is the Janus operator \eq{eq:weyl_poisson_braket} and
\begin{equation}
	\mc{I}_n \doteq \int \mathrm{d} \omega\,A(z)	\left( \overleftrightarrow{\mc{L}} \right)^n [\delta(D_{\rm H}) J].
\end{equation}
%

Now, let us calculate each of the terms appearing in \Eq{simp:eq:integral}. The $n=0$ term is simply given by
\begin{align}
	\mc{I}_0 
		&	=	\int \mathrm{d}\omega \, A(z) \delta(D_{\rm H}) J	
			=	A \boldsymbol{(} t,\vec{x},\Omega (t,\vec{x},\vec{k}) , \vec{k} \boldsymbol{)} \, 						J(t,\vec{x},\vec{k}).
\end{align}
For the $n=1$ term in \Eq{simp:eq:integral}, we write the Janus operator as $\overleftrightarrow{\mc{L}} \doteq \overleftarrow{\pd_\mu} \mc{J}^{\mu \nu} \overrightarrow{\pd_\nu}$, where $\mc{J}^{\mu \nu}$ is the canonical Poisson tensor in $z$ space. We then obtain
\begin{align}
	& \mc{I}_1 
			\doteq 	\int \mathrm{d}\omega \, (\pd_\mu A) \mc{J}^{\mu \nu} \pd_\nu[  \delta(D_{\rm H}) J	] \notag \\
		&	=	\int \mathrm{d}\omega \, \frac{\pd A}{\pd z^\mu}  \mc{J}^{\mu \nu} 
				\frac{\pd}{\pd z^\nu}\left[ \delta(\omega-\Omega)   J 	\right] \notag \\
		&	=	\int \mathrm{d}\omega \, \frac{\pd A}{\pd z^\mu} \mc{J}^{\mu \nu}  
				\left[ \frac{\pd \delta(\omega-\Omega)}{\pd z^\nu}   J	
				+ \delta(\omega-\Omega)  \frac{\pd J }{\pd z^\nu}  \right] \notag \\
		&	=	\int \mathrm{d}\omega \, \frac{\pd A}{\pd z^\mu} \mc{J}^{\mu \nu}  
				\left[ \delta'(\omega-\Omega) \frac{\pd ( \omega-\Omega)}{\pd z^\nu}   J
				+ \delta(\omega-\Omega)  \frac{\pd J }{\pd z^\nu}  \right] \notag \\
		&	=	\int \mathrm{d}\omega \,  
				\left[ - \frac{\pd }{\pd \omega} \left(  \frac{\pd A}{\pd z^\mu}   \mc{J}^{\mu \nu}  
				\frac{\pd ( \omega-\Omega)}{\pd z^\nu}  J \right) 
				+  \frac{\pd A}{\pd z^\mu} \mc{J}^{\mu \nu} \frac{\pd J }{\pd z^\nu}  \right]
				 \delta(\omega-\Omega)  \notag \\
		&	=	\left[ -   \frac{\pd^2 A}{\pd z^\mu \pd \omega}     \mc{J}^{\mu \nu}  
				\frac{\pd ( \omega-\Omega)}{\pd z^\nu}  J
				+  \frac{\pd A}{\pd z^\mu} \mc{J}^{\mu \nu} \frac{\pd J }{\pd z^\nu}  
				 \right]_{\omega=\Omega}.
	\label{simp:eq:calculation}
\end{align}
Here we integrated by parts in $\omega$ and used the fact that $J$ and $\Omega$ are independent of $\omega$. Explicitly writing the derivatives in terms of phase-space coordinates leads to
\begin{multline}
	\mc{I}_1
			=	 J \,
				\frac{\pd \Omega}{\pd t} \left( \frac{\pd^2 A}{\pd^2 \omega}  \right)_{w=\Omega}
				+ J \left( \frac{\pd^2 A}{ \pd t \pd \omega}  \right)_{w=\Omega}  \\
				+ J \left( \left\{\frac{\pd A}{\pd \omega},  \Omega\right\} 
							 \right)_{\omega=\Omega}  
		+  \left( \frac{\pd A }{\pd \omega}  \frac{\pd J}{\pd t} + \{ A, J\} \right)_{\omega=\Omega} ,
		\label{eq:integral_I}
\end{multline}
where $\{ \cdot , \cdot \} \doteq \overleftarrow{\pd_\vec{x}} \cdot \overrightarrow{\pd_\vec{k}} - \overleftarrow{\pd_\vec{k}} \cdot \overrightarrow{\pd_\vec{x}}$ is the canonical Poisson bracket.

After integrating by parts, we have been able to express $\mc{I}_1$ in \Eq{eq:integral_I} as a sum of derivatives acting on smooth functions. Since no singularities are present, then $\mc{I}_1 = O(\ep^0)$. By reiterating the same calculation in \Eq{simp:eq:calculation}, one can show that the higher $\mc{I}_n$ terms in \Eq{simp:eq:integral} are also smooth. Thus, one can approximate integrals as in \Eq{simp:eq:integral_orig} by their leading-order expansion in $\ep$. With these results, one can simplify the Moyal products and brackets in \Eq{eq:wke_wigner_moyal} in order to obtain the WKE \eq{eq:wke}.

In particular, let us calculate the special case where $A = D_H=\omega - \Omega$. Upon using \Eq{eq:integral_I}, we find
\begin{equation}
	\int  \mathrm{d}\omega \, D_{\rm H} \overleftrightarrow{\mc{L}} [ \delta(D_{\rm H}) J	] 
			=	 \pd_t J +  \{ J , \Omega \}  ,
	\label{simp:eq:integral_I_example}
\end{equation}
which is the advection term appearing in the WKE \eq{eq:wke}.

\subsection{Calculation of the nonlinear dissipation rate and the nonlinear source term}
\label{app:nonlinear}

We begin by calculating the Weyl symbol $F(t,\vec{x},\omega,\vec{k})$ corresponding to $\oper{F}$ in \Eq{eq:closure_F}. Upon substituting \Eq{eq:abstract_K}, we first note that the trace operator appearing in \Eq{eq:closure_F} can be written as
\begin{align}
	\mathrm{Tr}[ \,  \oper{K}(\msf{x}) \oper{W}  \oper{K}^\dag (\msf{y}) \oper{W} \, ] 
		=		&	\quad	\inner{\msf{x} \mid \oper{R}_j \,  \oper{W}	 \, \oper{L}_k^\dag		\mid \msf{y} } 
			\inner{\msf{y} \mid \oper{R}_k^\dag \, \oper{W} \, \oper{L}_j \mid \msf{x} }		 \notag \\
				& 	 
		+	\inner{\msf{x} \mid \oper{R}_j \, \oper{W} \, \oper{R}_k^\dag 	\mid \msf{y} }
			\inner{\msf{y} \mid \oper{L}_k^\dag \,	\oper{W} \,	\oper{L}	_j \mid \msf{x} }	
			 \notag \\
				& 		
		+	\inner{\msf{x} \mid \oper{L}_j \, \oper{W} \, \oper{L}_k^\dag   \mid \msf{y} }  
			\inner{\msf{y}\mid \oper{R}_k^\dag \,  \oper{W}\,  \oper{R}_j  \mid \msf{x} }		 \notag \\
				& 	
		+	\inner{\msf{x} \mid \oper{L}_j \, \oper{W} \, \oper{R}_k^\dag  \mid \msf{y} } 
			\inner{\msf{y} \mid \oper{L}_k^\dag \, \oper{W} \, \oper{R}_j  \mid \msf{x} },
	\label{eq:nonlinear_trace}
\end{align}
where the operators $\oper{L}_j$ and $\oper{R}_j$ are defined in \Eq{eq:abstract_oper}.

After substituting \Eq{eq:nonlinear_trace} into \Eq{eq:closure_F} and applying the Weyl transform \eq{eq:weyl_weyl_symbol}, we obtain integrals of the form
\begin{equation}
	\int \mathrm{d}^3 \msf{s} \, 
				\ue^{\ui \msf{k} \cdot \msf{s} / \ep} \,	
				\inner{\msf{x}+\tfrac{1}{2}\msf{s} \mid \oper{A} \mid \msf{x}-\tfrac{1}{2}\msf{s}} 
				\inner{\msf{x}-\tfrac{1}{2}\msf{s} \mid \oper{B} \mid \msf{x}+\tfrac{1}{2}\msf{s}},
	\label{eq:nonlinear_identity}
\end{equation}
where $\oper{A}$ and $\oper{B}$ represent the terms appearing in \Eq{eq:nonlinear_trace}. To evaluate these integrals, we shall use the identity
\begin{multline}
	\int \mathrm{d}^3 \msf{s} \, 
				\ue^{\ui \msf{k} \cdot \msf{s} / \ep} \,	
				\inner{\msf{x}+\tfrac{1}{2}\msf{s} \mid \oper{A} \mid \msf{x}-\tfrac{1}{2}\msf{s}} 
				\inner{\msf{x}-\tfrac{1}{2}\msf{s} \mid \oper{B} \mid \msf{x}+\tfrac{1}{2}\msf{s}} \\
		=	\int \frac{\mathrm{d}^3 \msf{p}  \, \mathrm{d}^3 \msf{q}	  }{(2\upi \ep )^3} 
			\delta^3(\msf{k}-\msf{p}-\msf{q})
			A(\msf{x},\msf{p} ) B(\msf{x},-\msf{q}).
\end{multline}
(This property is analogous to the convolution theorem frequently used in the Fourier transform.) Hence, we have
\begin{align}
	F(\msf{x}, \msf{k})	 =		\frac{1}{2}		
						\int \frac{\mathrm{d}^3 \msf{p}  \, \mathrm{d}^3 \msf{q}	  }{(2\upi  \ep)^3} 
						\delta^3(\msf{k}-\msf{p}-\msf{q})
							\big\{ & \quad
							   (R_j \star W \star L_k^*)(\msf{x},\msf{p}) (R_k^* \star W \star L_j )(\msf{x},-\msf{q}) 	\notag \\
				& 			+ (R_j \star W \star R_k^*)(\msf{x},\msf{p}) ( L_k^* \star W \star L_j ) (\msf{x},-\msf{q}) 	\notag \\
				& 			+ (L_j \star W \star L_k^*) (\msf{x},\msf{p})  (R_k^* \star W \star R_j )(\msf{x},-\msf{q})		\notag \\
				& 			+ ( L_j \star W \star R_k^*)(\msf{x},\msf{p}) ( L_k^* \star W \star R_j )(\msf{x},-\msf{q})  \, \, \, \,  \big\},
	\label{eq:nonlinear_F_aux}
\end{align}							 
where we used the Moyal product \eq{eq:weyl_Moyal}. Also, $L_j(\vec{k})=-(\vec{e}_z \times \vec{k})_j k_D^{-2}$ and $R_j(\vec{k}) = \vec{k}_j$ are respectively the Weyl symbols corresponding to the operators $\smash{\oper{L}_j}$ and $\smash{\oper{R}_j}$ in \Eqs{eq:abstract_oper}. The wave--wave nonlinearities are considered to be weak; hence, the Moyal products in \Eq{eq:nonlinear_F_aux} can be replaced by ordinary products. Hence,
\begin{equation}
	F(\msf{x}, \msf{k})		=		\frac{1}{2}	
							\int \frac{\mathrm{d}^3 \msf{p}  \, \mathrm{d}^3 \msf{q}	  }{(2\upi  \ep)^3} 
							\delta^3(\msf{k}-\msf{p}-\msf{q})
						    |M(\vec{p},\vec{q})|^2 W(\msf{x}, \msf{p} ) W(\msf{x},\msf{q}) 
							\, [ 1 + O(\ep) ] ,
	\label{eq:nonlinear_F_aux_II}
\end{equation}
where we used the reality property of $\fluctpsi$ so $W(\msf{x}, \msf{q}) = W(\msf{x},- \msf{q})$ and introduced $M(\vec{p},\vec{q}) \doteq	L_j(\vec{p}) R_j(\vec{q}) + L_k(\vec{q}) R_k(\vec{p})=	\vec{e}_z \cdot (\vec{p}\times \vec{q}) \left(  q_{\rm D}^{-2}  - p_{\rm D}^{-2} \right)$. [The Wigner function $W$ is a function of $(t,y,\omega,\vec{k})$ only. However, to simplify our notation, we used the arguments $(\msf{x},\msf{k})$ to denote the dependence on the phase-space variables.] Substituting the GO ansatz in \Eq{eq:wke_ansatz} into \Eq{eq:nonlinear_F_aux_II} and integrating in the frequency variables leads to
\begin{multline}
	F(t,y,\omega,\vec{k})	
		=	\ep \int \frac{\mathrm{d}^2\vec{p}  \, \mathrm{d}^2 \vec{q}	  }{(2\upi \ep )^2} 
							\delta^2(\vec{k}-\vec{p}-\vec{q}) \,
							\upi \delta\boldsymbol{(} 
									\omega
									- \Omega(t,\vec{x},\vec{p}) 
									- \Omega(t,\vec{x},\vec{q})
							 \boldsymbol{)}		\\ \times
							|M(\vec{p},\vec{q})|^2 
							J(t,y,\vec{p}) J(t,y,\vec{q})
							\, [ 1 + O(\ep) ],
	\label{eq:nonlinear_F}
\end{multline}
Substituting \Eq{eq:nonlinear_F} into \Eq{eq:wke_aux} and integrating in $\omega$ leads to the term $S_{\rm nl}(t, y,\vec{k})$ reported in \Eq{eq:wke_source_nl}.

Now, let us compute the Weyl symbol $\eta(\msf{x},\msf{k})$ of $\oper{\eta}$ in \Eq{eq:closure_eta}. Substituting \Eq{eq:abstract_K} into \Eq{eq:closure_eta} leads to
\begin{align}
	\oper{\eta}
		=		- \int \mathrm{d}^3 \msf{x} \, & \mathrm{d}^3 \msf{y} 
					\ketx \bralong{\msf{y}}
					(\oper{D}^{-1})^\dag 
				 \big[  
				 \quad \oper{L}_j \ketx \brax \oper{R}_j \, 		
				 \oper{W} \, 
				 \oper{L}_k^\dag  \kety \bray \oper{R}_k^\dag	\notag \\
			&	+ \oper{L}_j \ketx \brax \oper{R}_j \, 
				\oper{W} \, 
				\oper{R}_k^\dag \kety \bray \oper{L}_k^\dag 
				+ 	\oper{R}_j \ketx \brax \oper{L}_j \,
				\oper{W} \, 
				\oper{L}_k^\dag \kety \bray \oper{R}_k^\dag	\notag \\
			&	+
				\oper{R}_j \ketx \brax \oper{L}_j \, 
				\oper{W} \, 
				\oper{R}_k^\dag \kety\bray \oper{L}_k^\dag 
					\quad	\big] .
	\label{eq:nonlinear_eta_aux}
\end{align}
To calculate the Weyl transform of the above, we shall use the following result:
\begin{align}
	\msf{W}  \bigg[ \int \mathrm{d}^3 \msf{u} \,  \mathrm{d}^3 \msf{v}  \, &
								\ketlong{ \msf{u} } 
								\inner{ \msf{v} \mid \oper{A} \mid \msf{u} } \inner{ \msf{u} \mid \oper{B} \mid \msf{v} } \bralong{\msf{v}} \oper{C} \bigg]	\notag \\
	&	=	\msf{W} \left[ \int \mathrm{d}^3 \msf{u} \,  \mathrm{d}^3 \msf{v}  \,
								\ketlong{ \msf{u} } 
								\inner{ \msf{v} \mid \oper{A} \mid \msf{u} } \inner{ \msf{u} \mid \oper{B} \mid \msf{v} } \bralong{\msf{v}} \right]
				\star C( \msf{x}, \msf{k} ) \notag \\
	&	=	\left[ \int	\mathrm{d}^3 \msf{s}  \, \ue^{\ui \msf{k}\cdot \msf{s} / \ep }
		 		\inner{\msf{x}-\tfrac{1}{2}\msf{s} \mid \oper{A} \mid \msf{x}+\tfrac{1}{2}\msf{s}} 
				\inner{\msf{x}+\tfrac{1}{2}\msf{s} \mid \oper{B} \mid \msf{x}-\tfrac{1}{2}\msf{s}}
				\right ] \star C(\msf{x}, \msf{k} ) 	\notag 	\\
	&	=	\left[ \int \frac{\mathrm{d}^3 \msf{p}  \, \mathrm{d}^3 \msf{q}	  }{(2\upi \ep )^3} 
				\delta^3(\msf{k}-\msf{p}-\msf{q}) 
				B( \msf{x}, \msf{p} ) A( \msf{x}, -\msf{q} )  \right]
				\star C( \msf{x}, \msf{k} )
				\notag  \\
	&	=	\int \frac{\mathrm{d}^3 \msf{p}  \, \mathrm{d}^3 \msf{q}	  }{(2\upi \ep )^3} 
				\delta^3(\msf{k}-\msf{p}-\msf{q}) 
				B( \msf{x}, \msf{p} ) A( \msf{x}, -\msf{q} )   C( \msf{x}, \msf{k} )
				\, [ 1 + O(\ep) ],
	\label{eq:nonlinear_identity_II}
\end{align}
where in the third line we substituted \Eq{eq:nonlinear_identity}. We then calculate $\eta(\msf{x},\msf{k})$ in \Eq{eq:nonlinear_eta_aux} by using \Eq{eq:nonlinear_identity_II}. Similarly as in \Eq{eq:nonlinear_F}, we later approximate the Moyal products by ordinary products. To leading order, we obtain
\begin{multline}
	\mathrm{Im} [ \eta( \msf{x},  \msf{k} ) ]
		=	-		\int \frac{\mathrm{d}^3 \msf{p}  \, \mathrm{d}^3 \msf{q}	  }{(2\upi \ep )^3} 
					\, \delta^3(\msf{k}-\msf{p}-\msf{q}) \,
					\mathrm{Im} \{ [D^{-1}]^*( \msf{x}, -\msf{q} ) \} \\
				\times
                	M(\vec{p},-\vec{q}) M^*(\vec{p},\vec{k})  
					W( \msf{x}, \msf{p} )
					\, [ 1 + O(\ep) ].
\end{multline}
From \Eq{eq:wke_D}, we approximate $\mathrm{Im} \{ [D^{-1}]^*( \msf{x}, -\msf{q} ) \} \simeq \upi \delta \boldsymbol{(} q_0 - \Omega(t,y,\vec{q}) \boldsymbol{)}$. When substituting $W(t,y,\omega,\vec{k}) = 2 \upi \ep \delta \boldsymbol{(} \omega - \Omega(t,y,\vec{k}) \boldsymbol{)} J(t,y,\vec{k})$ and $M(\vec{p},-\vec{q})=-M(\vec{p},\vec{q})$, one obtains the following:
\begin{multline}
	\mathrm{Im} [ \eta( \msf{x},  \msf{k} ) ]
		=			\int \frac{ \mathrm{d}^2 \vec{p} \, \mathrm{d}^2 \vec{q}}{(2 \upi \ep )^2} \, 
					\delta^2(\vec{k} -\vec{p} - \vec{q}) \,
					\upi \delta\boldsymbol{(} 
									\omega
									- \Omega(t,\vec{x},\vec{p}) 
									- \Omega(t,\vec{x},\vec{q})
							 \boldsymbol{)}		\\
				\times
					M(\vec{p},\vec{q}) M^*(\vec{p},\vec{k})   
					J(t,y,\vec{p})
					\, [ 1 + O(\ep) ]. 
	\label{eq:nonlinear_eta}
\end{multline}
Finally, substituting \Eq{eq:nonlinear_eta} into \Eq{eq:wke_aux} and integrating in $\omega$ leads to the term $\gamma_{\rm nl}(t, y,\vec{k})$ reported in \Eq{eq:wke_gamma_nl}.


\subsection{Conservation of the total enstrophy $\mc{Z}$ and the total energy $\mc{E}$}
\label{app:conservation}

The time derivatives of the total enstrophy $\mc{Z}$ and total energy $\mc{E}$ are
\begin{subequations}
\begin{align}
\frac{\mathrm{d}\mc{Z}}{\mathrm{d}t}
	 &  = \frac{\ep^2}{2} 
	    \int \frac{\mathrm{d} y \, \mathrm{d}^2\vec{k}}{(2\upi)^2} \,
			    \left(
			    \frac{2 k_x k_y}{ k_{\rm D}^4}\,U''' J 
			    + \{ \Omega, J \} + 2\Gamma J  
			    \right) 
	  + \frac{\ep^3}{2}
	    \int \frac{\mathrm{d} y \, \mathrm{d}^2\vec{k}}{(2\upi \ep)^2} \, C[J,J], \\
 \frac{\mathrm{d}\mc{E}}{\mathrm{d}t} 
		  & = \frac{\ep^2}{2}
				\int \frac{\mathrm{d} y \, \mathrm{d}^2\vec{k}}{(2\upi \ep)^2} \,
				     \frac{1}{k_{\rm D}^2}\left(
				    -\frac{2 k_x k_y}{ k_{\rm D}^2} \, U' J 
				    + \{ \Omega , J  \} 
				    + 2 \Gamma J 
				    \right)
				    + \frac{\ep^3}{2}
	 		\int \frac{\mathrm{d} y \, \mathrm{d}^2\vec{k}}{(2\upi \ep)^2} \,
	 		\frac{1}{k_{\rm D}^2 } C[J,J],
\end{align}
\end{subequations}
where we substituted \Eqs{eq:wke}, \eq{eq:wke_zonal}, and \eq{eq:conservation_defs}. As shown by \cite{Ruiz:2016gv}, the first integrals on the right-hand side of the equations above are zero. Hence, in order to show conservation of total enstrophy and energy, one needs to show that
\begin{equation}
	\mc{G}(t) \doteq 
					\int \frac{\mathrm{d} y \, \mathrm{d}^2\vec{k}}{(2\upi \ep)^2} \, 
					\sigma(\vec{k}) \, C[J,J] =0,
\end{equation}
where $\sigma(\vec{k})=1$ for the case of enstrophy and $\sigma(\vec{k})=k_{\rm D}^{-2}$ for the case of energy. Since $C[J,J]$ is of the canonical form of the scattering operators found in homogeneous-turbulence theories, the proof of its conservation properties closely follows that sketched in \S 4.2.4 of \citet{Krommes:2002hva}. Substituting \Eq{eq:wke_scattering} leads to
\begin{multline}
	\mc{G}		=	\int \frac{	\mathrm{d} y \, \mathrm{d}^2\vec{k} \,
									\mathrm{d}^2\vec{p} \mathrm{d}^2\vec{q}}{(2\upi \ep)^4} \, 
						\delta^2(\vec{k} -\vec{p} -\vec{q}) \,
						\Theta(t,y,\vec{k},\vec{p},\vec{q})	 \, 
						\sigma(\vec{k})
						\big[	M(\vec{p}, \vec{q}) J(t,y,\vec{q})	\\
						- 2 	M(\vec{p}, \vec{k}) J(t,y,\vec{k})
						\big] M(\vec{p}, \vec{q})  J(t,y,\vec{p}).
\end{multline}
One can then exchange the momentum variables since they are simply integration variables. After using the symmetry property of $M(\vec{p},\vec{q})$ and the identity $\Theta(t,y,\vec{q},-\vec{p},\vec{k}) = \Theta(t,y,\vec{k},\vec{p},\vec{q})$, one can rewrite $\mc{G}$ as follows:
\begin{equation}
	\mc{G}		=	\int \frac{	\mathrm{d} y \, \mathrm{d}^2\vec{k} \,
									\mathrm{d}^2\vec{p} \mathrm{d}^2\vec{q}}{(2\upi \ep)^4} \, 
						\delta^2(\vec{k} -\vec{p} -\vec{q}) \,
						\Theta(t,y,\vec{k},\vec{p},\vec{q})	 \, 
						\chi(\vec{k},\vec{p},\vec{q}) \,  
						M(\vec{p}, \vec{q}) \, J(t,y,\vec{p}) \, J(t,y,\vec{q}),
\end{equation}
where 
\begin{equation}
	\chi(\vec{k},\vec{p},\vec{q}) 
		\doteq \sigma(\vec{k}) M(\vec{p}, \vec{q}) 
					- \sigma(\vec{p}) M(\vec{q}, \vec{k}) 
					 - \sigma(\vec{q}) M(\vec{k}, \vec{p}).
\end{equation} 
When substituting the spatial-resonance condition $\vec{k}=\vec{p}+ \vec{q}$, one shows that $\chi(\vec{k},\vec{p},\vec{q}) =0$ for both $\sigma(\vec{k})=1$ and $\sigma(\vec{k})=k_{\rm D}^{-2}$. Hence, $\mc{G}=0$, which proves that \Eqs{eq:wke} and \eq{eq:wke_zonal} conserve total enstrophy $\mc{Z}$ and total energy $\mc{E}$.



\begin{thebibliography}{58}
\expandafter\ifx\csname natexlab\endcsname\relax\def\natexlab#1{#1}\fi
\def\au#1{#1} \def\ed#1{#1} \def\yr#1{#1}\def\at#1{#1}\def\jt#1{\textit{#1}}
  \def\bt#1{#1}\def\bvol#1{\textbf{#1}} \def\vol#1{#1} \def\pg#1{#1}
  \def\publ#1{#1}\def\arxiv#1{#1}\def\org#1{#1}\def\st#1{\textit{#1}}

\bibitem[Ait-Chaalal {\em et~al.\/}(2016)Ait-Chaalal, Schneider, Meyer \&
  Marston]{AitChaalal:2016jx}
{\sc \au{Ait-Chaalal, F}, \au{Schneider, T}, \au{Meyer, B} \& \au{Marston,
  J~B}} \yr{2016}  \at{{Cumulant expansions for atmospheric flows}}.  \jt{New
  J. Phys.}  \bvol{18},  \pg{025019}.

\bibitem[Baker~Jr.(1958)]{BakerJr:1958bo}
{\sc \au{Baker~Jr., G~A}} \yr{1958}  \at{{Formulation of quantum mechanics
  based on the quasi-probability distribution induced on phase space}}.
  \jt{Phys. Rev.}  \bvol{109},  \pg{2198--2206}.

\bibitem[Biglari {\em et~al.\/}(1990)Biglari, Diamond \& Terry]{Biglari:1990hx}
{\sc \au{Biglari, H}, \au{Diamond, P~H} \& \au{Terry, P~W}} \yr{1990}
  \at{{Influence of sheared poloidal rotation on edge turbulence}}.  \jt{Phys.
  Fluids B}  \bvol{2},  \pg{1--4}.

\bibitem[Connaughton {\em et~al.\/}(2015)Connaughton, Nazarenko \&
  Quinn]{Connaughton:2015kk}
{\sc \au{Connaughton, Colm}, \au{Nazarenko, S} \& \au{Quinn, Brenda}} \yr{2015}
   \at{{Rossby and drift wave turbulence and zonal flows: The
  Charney--Hasegawa--Mima model and its extensions}}.  \jt{Phys. Rep.}
  \bvol{604},  \pg{1--71}.

\bibitem[Conway {\em et~al.\/}(2005)Conway, Scott, Schirmer, Reich, Kendl \&
  Team]{Conway:2005gq}
{\sc \au{Conway, G~D}, \au{Scott, B}, \au{Schirmer, J}, \au{Reich, M},
  \au{Kendl, A} \& \au{Team, the ASDEX~Upgrade}} \yr{2005}  \at{{Direct
  measurement of zonal flows and geodesic acoustic mode oscillations in ASDEX
  Upgrade using Doppler reflectometry}}.  \jt{Plasma Phys. Control. Fusion}
  \bvol{47},  \pg{1165--1185}.

\bibitem[Diamond {\em et~al.\/}(2005)Diamond, Itoh, Itoh \&
  Hahm]{Diamond:2005br}
{\sc \au{Diamond, P~H}, \au{Itoh, S-I}, \au{Itoh, K} \& \au{Hahm, T~S}}
  \yr{2005}  \at{{Zonal flows in plasma{\textemdash}a review}}.  \jt{Plasma
  Phys. Control. Fusion}  \bvol{47},  \pg{R35--R161}.

\bibitem[Diamond {\em et~al.\/}(1994)Diamond, Liang, Carreras \&
  Terry]{Diamond:1994fd}
{\sc \au{Diamond, P~H}, \au{Liang, Y~M}, \au{Carreras, B~A} \& \au{Terry, P~W}}
  \yr{1994}  \at{{Self-regulating shear flow turbulence: A paradigm for the $L$
  to $H$ transition}}.  \jt{Phys. Rev. Lett.}  \bvol{72},  \pg{2565--2568}.

\bibitem[Dodin(2014)]{Dodin:2014hw}
{\sc \au{Dodin, I~Y}} \yr{2014}  \at{{Geometric view on noneikonal waves}}.
  \jt{Phys. Lett. A}  \bvol{378},  \pg{1598--1621}.

\bibitem[Dorland {\em et~al.\/}(2000)Dorland, Jenko, Kotschenreuther \&
  Rogers]{Dorland:2000bb}
{\sc \au{Dorland, W}, \au{Jenko, F}, \au{Kotschenreuther, M} \& \au{Rogers,
  B~N}} \yr{2000}  \at{{Electron temperature gradient turbulence}}.  \jt{Phys.
  Rev. Lett.}  \bvol{85},  \pg{5579--5582}.

\bibitem[Farrell \& Ioannou(2003)]{Farrell:2003dm}
{\sc \au{Farrell, B~F} \& \au{Ioannou, P~J}} \yr{2003}  \at{{Structural
  stability of turbulent jets}}.  \jt{J. Atmos. Sci.}  \bvol{60},
  \pg{2101--2118}.

\bibitem[Farrell \& Ioannou(2007)]{Farrell:2007fq}
{\sc \au{Farrell, B~F} \& \au{Ioannou, P~J}} \yr{2007}  \at{{Structure and
  spacing of jets in barotropic turbulence}}.  \jt{J. Atmos. Sci.}  \bvol{64},
  \pg{3652--3665}.

\bibitem[Farrell \& Ioannou(2009)]{Farrell:2009ke}
{\sc \au{Farrell, B~F} \& \au{Ioannou, P~J}} \yr{2009}  \at{{A stochastic
  structural stability theory model of the drift wave-zonal flow system}}.
  \jt{Phys. Plasmas}  \bvol{16},  \pg{112903}.

\bibitem[Frisch \& Kolmogorov(1995)]{frisch1995turbulence}
{\sc \au{Frisch, U} \& \au{Kolmogorov, A~N}} \yr{1995} {\em {Turbulence: The
  Legacy of A. N. Kolmogorov}\/}.  \publ{Cambridge University Press}.

\bibitem[Fujisawa(2009)]{Fujisawa:2009jc}
{\sc \au{Fujisawa, A}} \yr{2009}  \at{{A review of zonal flow experiments}}.
  \jt{Nucl. Fusion}  \bvol{49},  \pg{013001}.

\bibitem[Hillesheim {\em et~al.\/}(2016)Hillesheim, Delabie, Meyer, Maggi,
  Meneses, Poli \& {JET Contributors}]{EUROfusionConsortium:2016bk}
{\sc \au{Hillesheim, J~C}, \au{Delabie, E}, \au{Meyer, H}, \au{Maggi, C~F},
  \au{Meneses, L}, \au{Poli, E} \& \au{{JET Contributors}}} \yr{2016}
  \at{{Stationary zonal flows during the formation of the edge transport
  barrier in the jet tokamak}}.  \jt{Phys. Rev. Lett.}  \bvol{116},
  \pg{065002}.

\bibitem[Imre {\em et~al.\/}(1967)Imre, {\"O}zizmir, Rosenbaum \&
  Zweifel]{Imre:1967fr}
{\sc \au{Imre, K}, \au{{\"O}zizmir, E}, \au{Rosenbaum, M} \& \au{Zweifel, P~F}}
  \yr{1967}  \at{{Wigner method in quantum statistical mechanics}}.  \jt{J.
  Math. Phys.}  \bvol{8},  \pg{1097--1108}.

\bibitem[Jenko {\em et~al.\/}(2000)Jenko, Dorland, Kotschenreuther \&
  Rogers]{Jenko:2000gn}
{\sc \au{Jenko, F}, \au{Dorland, W}, \au{Kotschenreuther, M} \& \au{Rogers,
  B~N}} \yr{2000}  \at{{Electron temperature gradient driven turbulence}}.
  \jt{Phys. Plasmas}  \bvol{7},  \pg{1904--1910}.

\bibitem[Kaw {\em et~al.\/}(2002)Kaw, Singh \& Diamond]{Kaw:2002ku}
{\sc \au{Kaw, P}, \au{Singh, R} \& \au{Diamond, P~H}} \yr{2002}  \at{{Coherent
  nonlinear structures of drift wave turbulence modulated by zonal flows}}.
  \jt{Plasma Phys. Control. Fusion}  \bvol{44},  \pg{51--59}.

\bibitem[Kim \& Diamond(2003)]{Kim:2003jf}
{\sc \au{Kim, Eun-jin} \& \au{Diamond, P~H}} \yr{2003}  \at{{Zonal flows and
  transient dynamics of the $L-H$ transition}}.  \jt{Phys. Rev. Lett.}
  \bvol{90},  \pg{185006}.

\bibitem[Kraichnan(2013)]{kraichnan2013closure}
{\sc \au{Kraichnan, R~H}} \yr{2013} {\em {The Closure Problem of Turbulence
  Theory}\/}.  \publ{Hardpress}.

\bibitem[Krommes(2002)]{Krommes:2002hva}
{\sc \au{Krommes, J~A}} \yr{2002}  \at{{Fundamental statistical descriptions of
  plasma turbulence in magnetic fields}}.  \jt{Phys. Rep.}  \bvol{360},
  \pg{1--352}.

\bibitem[Krommes \& Kim(2000)]{Krommes:2000ec}
{\sc \au{Krommes, J~A} \& \au{Kim, Chang-Bae}} \yr{2000}  \at{{Interactions of
  disparate scales in drift-wave turbulence}}.  \jt{Phys. Rev. E}  \bvol{62},
  \pg{8508--8539}.

\bibitem[Kuo(1949)]{Kuo:1949cr}
{\sc \au{Kuo, Hsiao-lan}} \yr{1949}  \at{{Dynamic instability of
  two-dimensional nondivergent flow in a barotropic atmosphere}}.  \jt{J.
  Meteor.}  \bvol{6},  \pg{105--122}.

\bibitem[Leslie(1973)]{Leslie:1973kv}
{\sc \au{Leslie, D~C}} \yr{1973}  \at{{Review of developments in turbulence
  theory}}.  \jt{Rep. Prog. Phys.}  \bvol{36},  \pg{1365--1424}.

\bibitem[Lin(1998)]{Lin:1998je}
{\sc \au{Lin, Z}} \yr{1998}  \at{{Turbulent transport reduction by zonal flows:
  massively parallel simulations}}.  \jt{Science}  \bvol{281},
  \pg{1835--1837}.

\bibitem[Littlejohn \& Winston(1993)]{Littlejohn:1993bd}
{\sc \au{Littlejohn, R~G} \& \au{Winston, Roland}} \yr{1993}  \at{{Corrections
  to classical radiometry}}.  \jt{J. Opt. Soc. Am. A}  \bvol{10},
  \pg{2024--2037}.

\bibitem[Malkov \& Diamond(2001)]{Malkov:2001kp}
{\sc \au{Malkov, M~A} \& \au{Diamond, P~H}} \yr{2001}  \at{{Bifurcation and
  scaling of drift wave turbulence intensity with collisional zonal flow
  damping}}.  \jt{Phys. Plasmas}  \bvol{8},  \pg{3996--4009}.

\bibitem[Malkov {\em et~al.\/}(2001)Malkov, Diamond \&
  Rosenbluth]{Malkov:2001hv}
{\sc \au{Malkov, M~A}, \au{Diamond, P~H} \& \au{Rosenbluth, M~N}} \yr{2001}
  \at{{On the nature of bursting in transport and turbulence in drift
  wave{\textendash}zonal flow systems}}.  \jt{Phys. Plasmas}  \bvol{8},
  \pg{5073--5076}.

\bibitem[Marston {\em et~al.\/}(2016)Marston, Chini \& Tobias]{Marston:2016ff}
{\sc \au{Marston, J~B}, \au{Chini, G~P} \& \au{Tobias, S~M}} \yr{2016}
  \at{{Generalized quasilinear approximation: Application to zonal jets}}.
  \jt{Phys. Rev. Lett.}  \bvol{116},  \pg{214501}.

\bibitem[Marston {\em et~al.\/}(2008)Marston, Conover \&
  Schneider]{Marston:2008gx}
{\sc \au{Marston, J~B}, \au{Conover, E} \& \au{Schneider, T}} \yr{2008}
  \at{{Statistics of an unstable barotropic jet from a cumulant expansion}}.
  \jt{J. Atmos. Sci.}  \bvol{65},  \pg{1955--1966}.

\bibitem[McDonald(1988)]{McDonald:1988dp}
{\sc \au{McDonald, S~W}} \yr{1988}  \at{{Phase-space representations of wave
  equations with applications to the eikonal approximation for short-wavelength
  waves}}.  \jt{Phys. Rep.}  \bvol{158},  \pg{337--416}.

\bibitem[McDonald(1991)]{McDonald:1991kk}
{\sc \au{McDonald, S~W}} \yr{1991}  \at{{Wave kinetic equation in a fluctuating
  medium}}.  \jt{Phys. Rev. A}  \bvol{43},  \pg{4484--4499}.

\bibitem[McDonald \& Kaufman(1985)]{McDonald:1985ib}
{\sc \au{McDonald, S~W} \& \au{Kaufman, A~N}} \yr{1985}  \at{{Weyl
  representation for electromagnetic waves: The wave kinetic equation}}.
  \jt{Phys. Rev. A}  \bvol{32},  \pg{1708--1713}.

\bibitem[Moyal(1949)]{Moyal:1949gj}
{\sc \au{Moyal, J~E}} \yr{1949}  \at{{Quantum mechanics as a statistical
  theory}}.  \jt{Math. Proc. Cambridge Philos. Soc.}  \bvol{45},  \pg{99--124}.

\bibitem[Nazarenko(2011)]{Nazarenko:2011cr}
{\sc \au{Nazarenko, S~V}} \yr{2011} {\em {Wave Turbulence}\/}, 1st edn.,
  \st{Lecture Notes in Physics},  \vol{vol. 825}.  \publ{New York: Springer}.

\bibitem[Numata {\em et~al.\/}(2007)Numata, Ball \& Dewar]{Numata:2007ek}
{\sc \au{Numata, Ryusuke}, \au{Ball, Rowena} \& \au{Dewar, R~L}} \yr{2007}
  \at{{Bifurcation in electrostatic resistive drift wave turbulence}}.
  \jt{Phys. Plasmas}  \bvol{14},  \pg{102312}.

\bibitem[Ogura(1963)]{Ogura:2006ck}
{\sc \au{Ogura, Y}} \yr{1963}  \at{{A consequence of the zero-fourth-cumulant
  approximation in the decay of isotropic turbulence}}.  \jt{J. Fluid Mech.}
  \bvol{16},  \pg{33--40}.

\bibitem[Parker(2014)]{Parker:2014tb}
{\sc \au{Parker, J~B}} \yr{2014}  \at{{Zonal Flows and Turbulence in Fluids and
  Plasmas}}. PhD thesis, Princeton University.

\bibitem[Parker(2016)]{Parker:2016eu}
{\sc \au{Parker, J~B}} \yr{2016}  \at{{Dynamics of zonal flows: failure of
  wave-kinetic theory, and new geometrical optics approximations}}.  \jt{J.
  Plasma Phys.}  \bvol{82},  \pg{595820602}.

\bibitem[Parker(2018)]{Parker:2018eda}
{\sc \au{Parker, J~B}} \yr{2018}  \at{{Numerical simulation of the
  geometrical-optics reduction of CE2 and comparisons to quasilinear
  dynamics}}.  \jt{Phys. Plasmas}  \bvol{25},  \pg{055708}.

\bibitem[Parker \& Krommes(2013)]{Parker:2013hy}
{\sc \au{Parker, J~B} \& \au{Krommes, J~A}} \yr{2013}  \at{{Zonal flow as
  pattern formation}}.  \jt{Phys. Plasmas}  \bvol{20},  \pg{100703}.

\bibitem[Parker \& Krommes(2014)]{Parker:2014fc}
{\sc \au{Parker, J~B} \& \au{Krommes, J~A}} \yr{2014}  \at{{Generation of zonal
  flows through symmetry breaking of statistical homogeneity}}.  \jt{New J.
  Phys.}  \bvol{16},  \pg{035006}.

\bibitem[Peskin \& Schroeder(1995)]{Peskin:2018}
{\sc \au{Peskin, M~E} \& \au{Schroeder, D~V}} \yr{1995} {\em {An Introduction
  to Quantum Field Theory}\/}.  \publ{Westview Press}.

\bibitem[Ruiz(2017)]{Ruiz:2017wc}
{\sc \au{Ruiz, D~E}} \yr{2017}  \at{{A Geometric Theory of Waves and Its
  Applications to Plasma Physics}}. PhD thesis, Princeton University,
  Princeton.

\bibitem[Ruiz {\em et~al.\/}(2016)Ruiz, Parker, Shi \& Dodin]{Ruiz:2016gv}
{\sc \au{Ruiz, D~E}, \au{Parker, J~B}, \au{Shi, E~L} \& \au{Dodin, I~Y}}
  \yr{2016}  \at{{Zonal-flow dynamics from a phase-space perspective}}.
  \jt{Phys. Plasmas}  \bvol{23},  \pg{122304}.

\bibitem[Singh {\em et~al.\/}(2014)Singh, Singh, Kaw, G{\"u}rcan \&
  Diamond]{Singh:2014bh}
{\sc \au{Singh, Rameswar}, \au{Singh, R}, \au{Kaw, P}, \au{G{\"u}rcan, {\"O}~D}
  \& \au{Diamond, P~H}} \yr{2014}  \at{{Coherent structures in ion temperature
  gradient turbulence-zonal flow}}.  \jt{Phys. Plasmas}  \bvol{21},
  \pg{102306}.

\bibitem[Smolyakov \& Diamond(1999)]{Smolyakov:1999jk}
{\sc \au{Smolyakov, A~I} \& \au{Diamond, P~H}} \yr{1999}  \at{{Generalized
  action invariants for drift waves-zonal flow systems}}.  \jt{Phys. Plasmas}
  \bvol{6},  \pg{4410--4413}.

\bibitem[Smolyakov {\em et~al.\/}(2000)Smolyakov, Diamond \&
  Malkov]{Smolyakov:2000be}
{\sc \au{Smolyakov, A~I}, \au{Diamond, P~H} \& \au{Malkov, M}} \yr{2000}
  \at{{Coherent structure phenomena in drift wave{\textendash}zonal flow
  turbulence}}.  \jt{Phys. Rev. Lett.}  \bvol{84},  \pg{491--494}.

\bibitem[Srinivasan \& Young(2012)]{Srinivasan:2012im}
{\sc \au{Srinivasan, K} \& \au{Young, W~R}} \yr{2012}  \at{{Zonostrophic
  instability}}.  \jt{J. Atmos. Sci.}  \bvol{69},  \pg{1633--1656}.

\bibitem[Tobias \& Marston(2013)]{Tobias:2013hk}
{\sc \au{Tobias, S~M} \& \au{Marston, J~B}} \yr{2013}  \at{{Direct statistical
  simulation of out-of-equilibrium jets}}.  \jt{Phys. Rev. Lett.}  \bvol{110},
  \pg{104502}.

\bibitem[Tobias \& Marston(2016)]{Tobias:2016kp}
{\sc \au{Tobias, S~M} \& \au{Marston, J~B}} \yr{2016}  \at{{Three-dimensional
  rotating Couette flow via the generalised quasilinear approximation}}.
  \jt{J. Fluid Mech.}  \bvol{810},  \pg{412--428}.

\bibitem[Tracy {\em et~al.\/}(2014)Tracy, Brizard, Richardson \&
  Kaufman]{Tracy:2014to}
{\sc \au{Tracy, E~R}, \au{Brizard, A~J}, \au{Richardson, A~S} \& \au{Kaufman,
  A~N}} \yr{2014} {\em {Ray Tracing and Beyond: Phase Space Methods in Plasma
  Wave Theory}\/}.  \publ{New York: Cambridge University Press}.

\bibitem[Trines {\em et~al.\/}(2005)Trines, Bingham, Silva, Mendon{\c c}a,
  Shukla \& Mori]{Trines:2005in}
{\sc \au{Trines, R}, \au{Bingham, R}, \au{Silva, L~O}, \au{Mendon{\c c}a, J~T},
  \au{Shukla, P~K} \& \au{Mori, W~B}} \yr{2005}  \at{{Quasiparticle approach to
  the modulational instability of drift waves coupling to zonal flows}}.
  \jt{Phys. Rev. Lett.}  \bvol{94},  \pg{165002}.

\bibitem[Weyl(1931)]{Weyl:1931uw}
{\sc \au{Weyl, H}} \yr{1931} {\em {The Theory of Groups and Quantum
  Mechanics}\/}.  \publ{New York: Dover}.

\bibitem[Wigner(1932)]{Wigner:1932cz}
{\sc \au{Wigner, E}} \yr{1932}  \at{{On the quantum correction for
  thermodynamic equilibrium}}.  \jt{Phys. Rev.}  \bvol{40},  \pg{749--759}.

\bibitem[Zhu {\em et~al.\/}(2018{\natexlab{{\em a\/}}})Zhu, Zhou \&
  Dodin]{Zhu:2018hk}
{\sc \au{Zhu, H}, \au{Zhou, Y} \& \au{Dodin, I~Y}} \yr{2018{\natexlab{{\em
  a\/}}}}  \at{{On the Rayleigh--Kuo criterion for the tertiary instability of
  zonal flows}}.  \jt{Phys. Plasmas}  \bvol{25},  \pg{082121}.

\bibitem[Zhu {\em et~al.\/}(2018{\natexlab{{\em b\/}}})Zhu, Zhou \&
  Dodin]{Zhu:2018gk}
{\sc \au{Zhu, H}, \au{Zhou, Y} \& \au{Dodin, I~Y}} \yr{2018{\natexlab{{\em
  b\/}}}}  \at{{On the structure of the drifton phase space and its relation to
  the Rayleigh{\textendash}Kuo criterion of the zonal-flow stability}}.
  \jt{Phys. Plasmas}  \bvol{25},  \pg{072121}.

\bibitem[Zhu {\em et~al.\/}(2018{\natexlab{{\em c\/}}})Zhu, Zhou, Ruiz \&
  Dodin]{Zhu:2018fd}
{\sc \au{Zhu, H}, \au{Zhou, Y}, \au{Ruiz, D~E} \& \au{Dodin, I~Y}}
  \yr{2018{\natexlab{{\em c\/}}}}  \at{{Wave kinetics of drift-wave turbulence
  and zonal flows beyond the ray approximation}}.  \jt{Phys. Rev. E}
  \bvol{97},  \pg{053210}.

\end{thebibliography}
\end{document}